\documentclass[aps,prd,twocolumn,showpacs]{revtex4-1}
\usepackage{graphicx}
\usepackage{amsmath,amssymb,latexsym}
\usepackage{bm}
\usepackage{float} 
\usepackage{aas_macros} 

\begin{document}

\title{Parametrization of the nucleus--nucleus gamma-ray production cross sections below 100~GeV/nucleon: Subthreshold pions and Hard photons}

\author{Ervin~Kafexhiu}
\affiliation{Max-Planck-Institut f\"ur Kernphysik, Saupfercheckweg 1, D-69117 Heidelberg, Germany}
\email[E-mail: ]{ervin.kafexhiu@mpi-hd.mpg.de}

\begin{abstract}
``Subthreshold pions'' and so-called ``hard photons'' are two important channels for producing less than 1~GeV $\gamma$-rays and $e^\pm$ pairs from nuclear collisions with energy per nucleon below the $\pi$-meson production threshold. I use publicly available experimental data to parametrize these two channels' $\gamma$-ray and $e^\pm$ production cross sections and extend the pion contribution to these particles spectra at higher energies using their corresponding spectra from $pp$ interactions. These parametrizations are valid for collision energy $T_p\leq 100$~A~GeV and agree reasonably well with the available experimental data. The new parametrizations allow, for the first time, accurate studies of astrophysical $\gamma$-rays below 1~GeV.
\end{abstract}


\maketitle

\section{Introduction \label{intro}}
Nuclear interactions are ubiquitous in the universe, ranging from thermal plasmas to high energy cosmic rays, being an abundant source of $\gamma$-rays and other secondary particles relevant to astrophysics. At low collision energies and above the Coulomb barrier, inelastic collisions and/or nuclear reactions produce excited nuclei in the final state. De-excitation of these nuclear levels can produce prompt $\gamma$-ray lines as well as a continuum component which is a superposition of many $\gamma$-ray lines with relatively large widths. The main emission lines appear between 0.1 and 10~MeV. At higher collision energies, say $T_p\gtrsim 10$~A~MeV (MeV/nucleon), nuclear collective modes are excited and produce an additional continuum $\gamma$-radiation component. The most prominent source of this continuum, is the so-called giant dipole resonance which emits most of the $\gamma$-rays between 10 and 25 MeV. The $\gamma$-rays with energy below 25-30 MeV are referred to as the \textit{statistical photons} and their origin is the nuclear structure. 

For nuclear collisions with energy above the $\pi$-meson production threshold ($T_p>T_{p\,NN}^{\rm th}\approx0.28$~A~GeV), pions are expected to be produced effectively. Nuclear interactions at such energies produce pions through individual nucleon--nucleon collisions. The decay of these pions produce $\gamma$-rays, $e^\pm$ pairs and neutrinos. Similar with $pp$ interactions, the main source of the $\gamma$-rays at high energy nuclear collisions is the decay of the $\pi^0$-meson and to a lesser extent the decay of the $\eta$-meson. 

These channels however, are not the only ones through which nuclear interactions can produce $\gamma$-rays. Experimental observations show that the continuum radiation from low energy nuclear interactions does not stop with the statistical photons. For photon energies above the giant dipole resonance the $\gamma$-ray spectrum changes its slope and becomes harder. This new direct $\gamma$-ray channel is called the \textit{hard photons}. Moreover, experimental observations show that nuclear collisions unlike nucleon--nucleon interactions, can produce pions at energies $T_p<T_{p\,NN}^{\rm th}$. These pions are called the \textit{subthreshold pions} and are observed for collision energies as low as 20~A~MeV \cite{Grimm1985, Grosse1986b}. From energetics point of view, the hard photons and subthreshold pions require a mechanism that will extract the energy from many nucleons inside the nucleus. They are partly explained by the Fermi motion, however, the detailed physics related to the cooperative effects is not yet well understood.

Experimental observations have played a crucial role in singling out the main processes responsible for the production of hard photons and subthreshold pions. For instance, the center of mass frame observations show that hard photons have a dipole angular distribution and their production source velocity is close to the nucleon--nucleon velocity in this frame. These suggest that the source of hard photons is the neutron--proton ($np$) bremsstrahlung that occurs during the early stage of the nuclear interaction \citep[see e.g.][]{Nifenecker1989a, Cassing1990,Schutz(1997)}. The $pp$ bremsstrahlung has a quadrupole nature, therefore, gives a minor contribution compared to the $np$. Moreover, it is observed that hard photon energy distribution is an exponential function of the form $\sim\exp(-E_\gamma/E_0^\gamma)$, with an inverse slope parameter $E_0^\gamma$ that is experimentally determined. In heavy ion experiments an additional component of direct hard photons is observed, called the \textit{thermal hard photons} \citep[see e.g.][]{DEnterria2000, DEnterria2001}. This component however, is not important for light nuclei that are relevant in astrophysics.

Subthreshold pions are also produced during the first stage of the nucleus--nucleus collision. In contrast to the hard photons, pion's mean free path in the nuclear matter is short. As a result, they are reabsorbed and re-emitted several times which leads to their thermalization with the nuclear matter. Therefore, pions energy distribution carries important information about the fireball that is formed in the intermediate phase of the nuclear reaction. Experiments show that the pion energy distribution can be fitted with a Maxwellian distribution. At energies $T_p\approx 1$~A~GeV, $\Delta$-resonance production becomes significant. Its decay results in the formation of a high energy tail on top of the Maxwellian distribution. At such energies, the experimental data are fitted with more than one Maxwellian distribution \citep[see e.g.][]{Stock1986, Cassing1990,Senger1999,Bonasera2006}.

Although for $T_p \lesssim T_{p\,NN}^{\rm th}$ interactions there is no theory that can accurately predict the hard photon and subthreshold pion production cross sections, at higher energies however, one can use Glauber's multiple scattering theory \citep{Glauber1955,Franco1966, Glauber1970} as applied in some superposition model (e.g. the \textit{wounded-nucleon model} \citep{Bialas1976} or the \textit{additive quark model} \citep[see e.g.][]{Bialas1977,Anisovich1978,Nikolaev1981}) to calculate the secondary particle production average and dispersion multiplicity distributions. In these models, the hadron--nucleus and nucleus--nucleus collision are treated as a sequence of nucleon--nucleon or quark(s)--nucleon scatterings. As a result, the average secondary particle production multiplicity of a nucleus--nucleus collision is proportional with the average yield produced by nucleon--nucleon or quark(s)--nucleon interactions. The proportionality factor is called the number of wounded constituents. The so-called \textit{nuclear enhancement factor} \citep[see e.g.][]{Gaisser1992}, is an application of the wounded nucleon model in astrophysics. This quantity sums the contributions of all nucleus--nucleus interactions which scale the secondary particle production spectrum produced by $pp$ collisions. 

By combining subthreshold pions at low energies with high energy pion production calculations, it is possible to compute pion production cross sections for a wide energy range which can be important in astrophysics. The hard photon and subthreshold pion channels allow nuclear interactions to produce $\gamma$-rays and $e^\pm$ pairs at low energies for which the $pp$ interactions do not. Moreover, the $\gamma$-rays from these two channels, significantly contribute in the $\gamma$-ray spectrum below the $\pi^0$-bump that is produced by $pp$ interactions. These channels should be taken into account in the identification of the radiation process and the parent particles that produce the $\gamma$-rays (leptonic versus hadronic). This identification should be based on the shape of the measured $\gamma$-ray spectrum below 1~GeV. Surprisingly, so far this question has not been studied even on a qualitative level despite the recent numerous claims of detection of hadronic $\gamma$-rays based on the observations by Fermi-LAT.

In this paper I intend to fill the gap of studies in this area. Although the physics of complex processes of nucleus--nucleus interactions at low energies is not yet fully understood and described by an adequate theory, the available experimental measurements are quite comprehensive to conduct a detailed quantitative study of this important issue. Note that the units used throughout this article are the natural units (i.e. $\hbar=c=k_B=1$).

\section{Hard photons production cross sections \label{sec:hardGam}}
The experimentally supported assertion that the direct photons with energy $E_\gamma>30$~MeV are produced through incoherent neutron--proton ($np$) bremsstrahlung is further supported by the fact that their production cross section scales with the number of first $np$ collisions. Following \citet{Bertholet1987}, the hard photons production cross section can be parametrized as follows:
\begin{equation}\label{eq:XSg}
\sigma_\gamma=\sigma_R\,\left\langle N_{np}\right\rangle_b\,P_\gamma,
\end{equation}

\noindent where $\sigma_R$ is the reaction cross section, $\left\langle N_{np}\right\rangle_b$ is the total number of initial $np$ collisions averaged over the impact parameter and $P_\gamma$ is the $\gamma$-ray emission probability in a single collision. We use the parametrization of $\sigma_R$ from \citet{Cassing1990} which has the form:
\begin{equation}\label{eq:XSr}
\sigma_R = 10\pi\,r_0^2\left(A_p^{1/3}+A_t^{1/3}+b \right)^2 \! \left(1-\frac{V_c}{A_p\,T_p}\right)~\![{\rm mb}].
\end{equation}
\noindent Here, $r_0=1.16$~fm, $b=2.0$ and $T_p$ is the projectile kinetic energy per nucleon in the laboratory frame. $V_c=1.44\,Z_p Z_t/R$ is the Coulomb potential of the colliding nuclei in MeV units and  $R=1.2\,(A_p^{1/3} + A_t^{1/3})$~fm.  $Z_p$, $A_p$ and $Z_t$ and $A_t$ are the charge and mass numbers for the projectile and target nuclei, respectively. The parameter $\left\langle N_{np}\right\rangle_b$ is given by:
\begin{equation}
\left\langle N_{np}\right\rangle_b=\left\langle A_F\right\rangle\,\frac{Z_p\,N_t+Z_t\,N_p}{A_p\,A_t},
\end{equation}
\noindent where $N_p$ and $N_t$ are the number of neutrons for the projectile and target, respectively. $\left\langle A_F\right\rangle$ is the number of nucleon-nucleon collisions which is given by:
\begin{equation}
\left\langle A_F\right\rangle = A_p\times\frac{5\,A_t^{2/3} - A_p^{2/3}}{5\left(A_p^{1/3}+A_t^{1/3}\right)^2},
\end{equation}
\noindent and it is valid for $A_p\leq A_t$ \cite{Nifenecker1985}.

Using experimental data for the hard photon production cross section $\sigma_\gamma$ for $E_\gamma>E_\gamma^{\rm min}=30$~MeV, \citet{Cassing1990} has parametrized the $\gamma$-ray emission probability $P_\gamma$ as: 
\begin{equation}\label{eq:Pgamma}
P_\gamma=M_0\times\exp\left(-\frac{E_\gamma^{\rm min}}{E_0^\gamma}\right),
\end{equation}
\noindent where, $M_0=(5.5\pm0.1)\times10^{-4}$ is a constant that is derived from fitting the cross section data; whereas, $E_0^\gamma$ is found from fitting the hard photon energy distribution spectrum, which has an exponential shape $d\sigma/dE_\gamma\sim \exp(-E_\gamma/E_0^\gamma)$. If we normalize this function such that its integral from $E_\gamma^{\rm min}$ to infinity gives $\sigma_\gamma$, the hard photon differential cross section becomes:
\begin{equation}\label{eq:dXSdEg0}
\frac{d\sigma_\gamma}{dE_\gamma}=\frac{\sigma_\gamma}{E_0^\gamma}\times\exp\left(\frac{E_\gamma^{\rm min} - E_\gamma}{E_0^\gamma}\right).
\end{equation}

\begin{figure}
\includegraphics[scale=0.47]{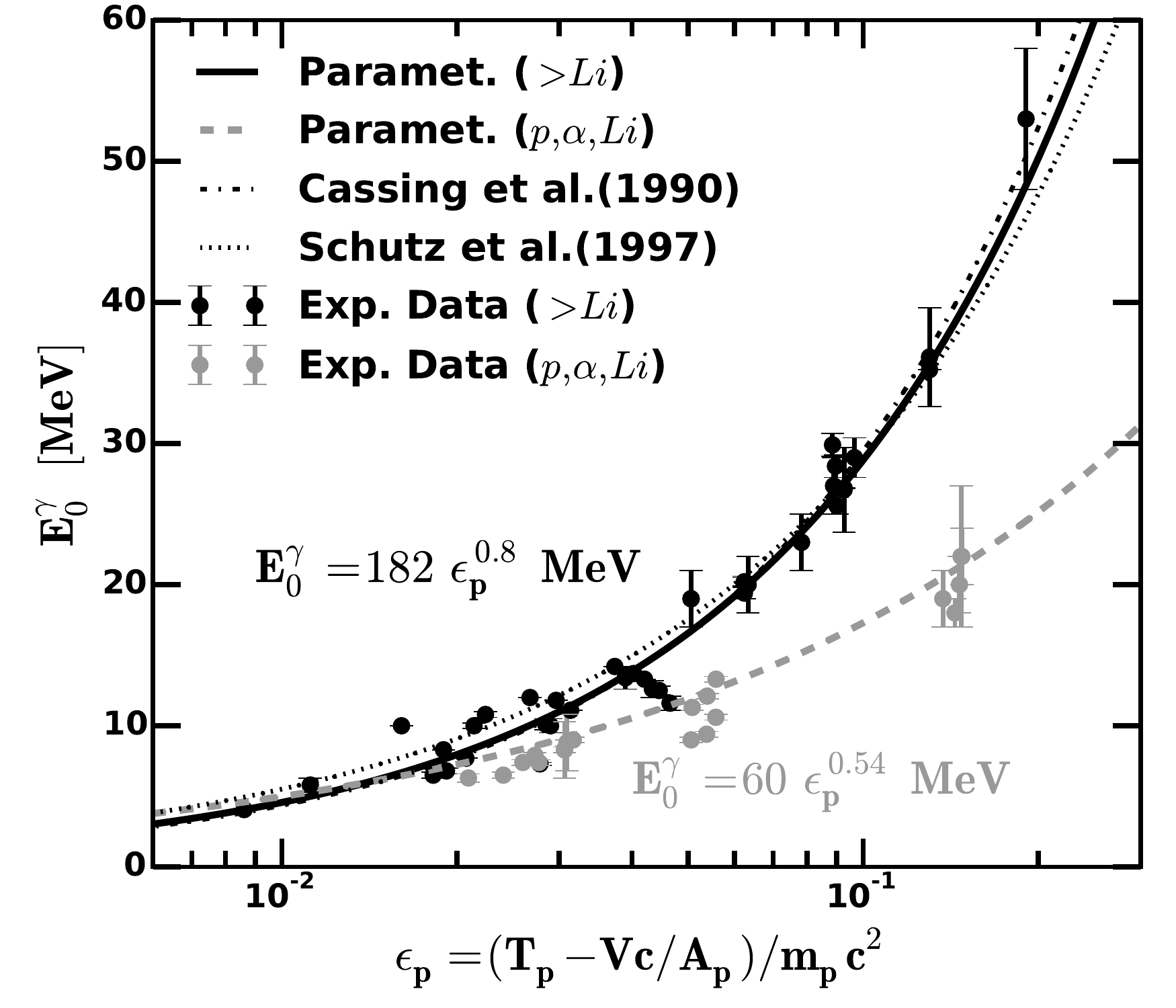}
\caption{Hard photon inverse slope parameter $E_0^\gamma$ for $E_\gamma>30$~MeV as a function of $\epsilon_p=\left(T_p-V_c/A_p\right)/m_p$, where, $T_p$ is the projectile kinetic energy per nucleon, $V_c$ is the Coulomb energy, $A_p$ the projectile mass number and $m_p$ is the nucleon mass in units of energy. The experimental data that are used here are listed in Table~\ref{tab:E0g}. Data in grey color correspond to light projectiles $p$, $\alpha$ and Li, whereas, data in black color belongs to heavier projectiles. The grey thick-dash-line shows the Eq.~(\ref{eq:E0g}) parametrization for projectiles $p$, $\alpha$ and Li which is $E_0^\gamma = (60\pm10)\,\epsilon_p^{0.54\pm0.06}$~MeV; whereas, the black thick-line shows the parametrization for projectiles heavier than Li $E_0^\gamma = (182\pm1)\,\epsilon_p^{0.805\pm0.002}$~MeV. The thin dash-dot line is the parametrization given in \citet{Cassing1990} and the dots curve is the parametrization given in \citet{Schutz(1997)}. \label{fig:E0g}}
\end{figure}

The hard photon inverse slope parameter $E_0^\gamma$ is the only missing element in Eqs.~(\ref{eq:Pgamma}) and (\ref{eq:dXSdEg0}). Experimental data show that $E_0^\gamma$ systematically increases with the collision energy and the available data show two different trends for heavy and for light ions interactions. I have compiled here publicly available experimental data on $E_0^\gamma$ that are found in the literature and they are recorded in Table~\ref{tab:E0g}. These data include a variety of colliding systems for a broad collision energy range. I have parametrized these data in two branches: Data for light projectiles $p$, $\alpha$ ($^4$He) and Li, and data for heavier projectiles. These two groups seem to form two distinct clusters and $E_0^\gamma$ for projectiles heavier than Li is larger than for lighter projectiles. This difference might be related to the absence of Fermi motion of nucleons inside the light nuclei which leads to lesser energetic $np$ collisions, thus to a softer spectrum of $\gamma$-rays. The parametrization of the inverse slope parameter has the form:
\begin{equation}\label{eq:E0g}
E_0^\gamma = a\,\epsilon_p^b,
\end{equation}

\noindent where $\epsilon_p=(T_p-V_c/A_p)/m_p$ is a dimensionless variable, and $m_p$ is the nucleon rest mass which is considered equal to the proton mass. By fitting the experimental data, I find that for light projectiles ($p$, $\alpha$ and Li) $a=60\pm10$~MeV and $b=0.54\pm0.06$ and for heavier projectiles $a=182\pm1$~MeV and $b=0.805\pm0.002$. Figure~\ref{fig:E0g} compares the experimental data with the parametrization described here as well as with the parametrizations described in \citep{Cassing1990, Schutz(1997)}. The figure shows that all parametrizations agree with each other and fit reasonably well the available experimental data.

\begin{figure}
\includegraphics[scale=0.47]{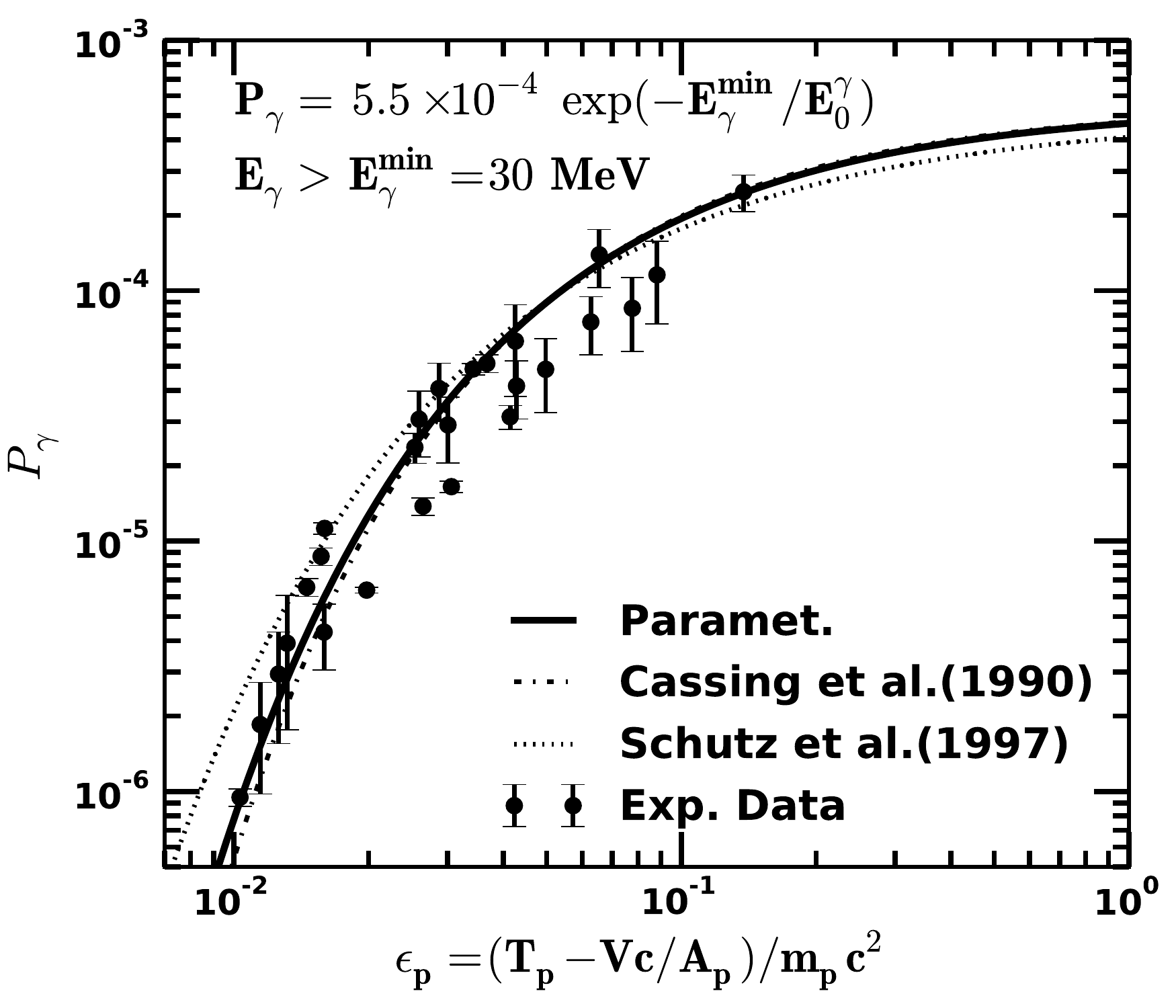}
\caption{Hard photon emission probability $P_\gamma$ for $E_\gamma>E_\gamma^{\rm min}=30$~MeV as a function of $\epsilon_p$. The experimental data points are compiled by \citet{Cassing1990}. The black line is the parametrization shown in Eqs.~(\ref{eq:Pgamma}) and (\ref{eq:E0g}). The thin dash-dot line and the dotted line are the parametrizations given in \citet{Cassing1990} and \citet{Schutz(1997)}, respectively. \label{fig:Pg}}
\end{figure}

Figure~\ref{fig:Pg} compares the $P_\gamma$ emission probability parametrization in Eq.~(\ref{eq:Pgamma}) using $E_0^\gamma$ from Eq.~(\ref{eq:E0g}), against the experimental data compiled in \citet{Cassing1990}. The parametrizations used in \citet{Cassing1990} and \citet{Schutz(1997)} are also included and, they differ only on the parametrization of $E_0^\gamma$. All these parametrizations agree reasonably well with each other and with the available experimental data.

If we include Eq.~(\ref{eq:XSg}) and Eq.~(\ref{eq:Pgamma}) in Eq.~(\ref{eq:dXSdEg0}), the hard photons differential cross section for $E_\gamma>E_\gamma^{\rm min}=30$~MeV is further simplified:
\begin{equation}\label{eq:BremsSpec}
\frac{d\sigma_\gamma}{dE_\gamma}=\frac{\sigma_R\,\left\langle N_{np}\right\rangle_b\,M_0}{E_0^\gamma}\times\exp\left(-\frac{E_\gamma}{E_0^\gamma}\right).
\end{equation}

\begin{figure}
\includegraphics[scale=0.45]{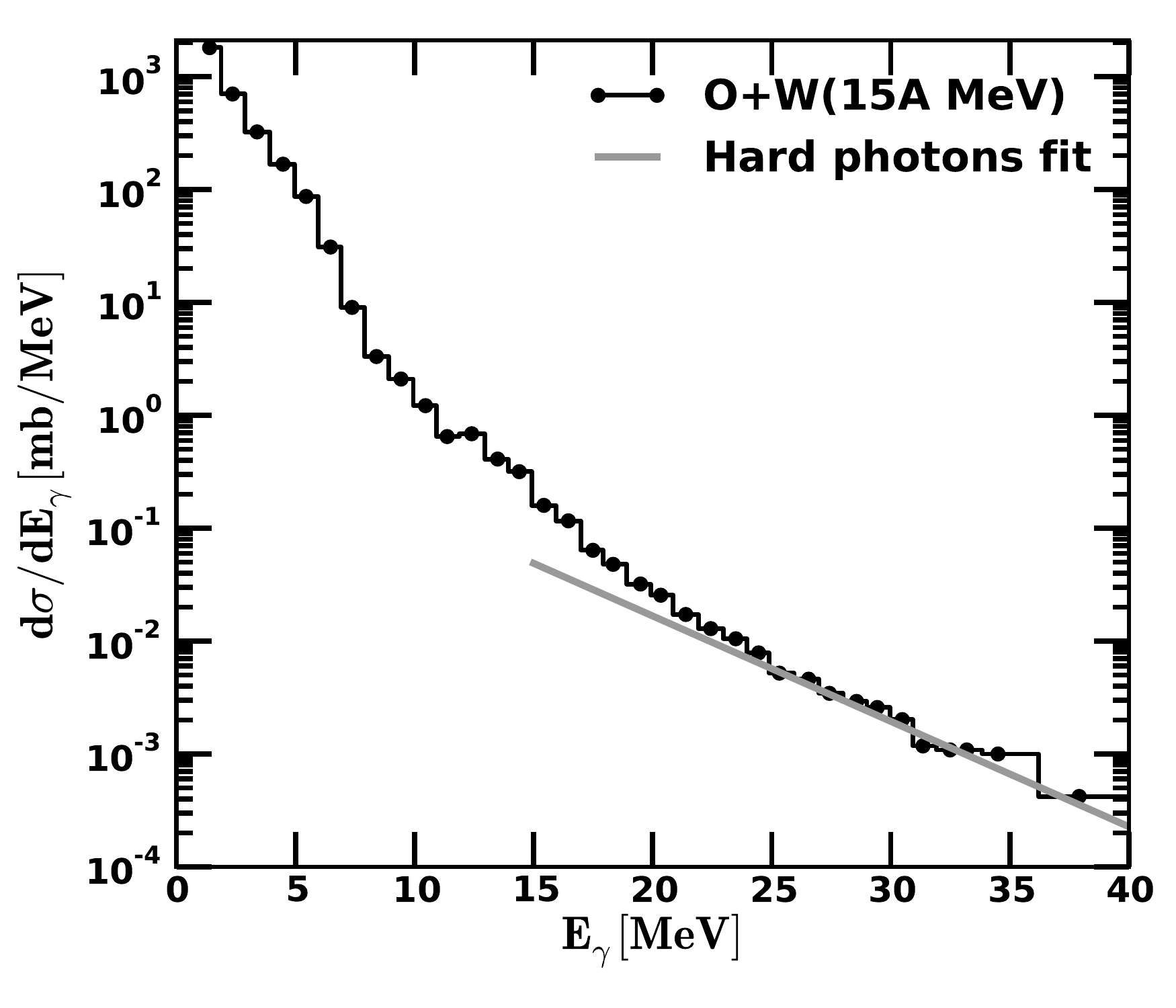}
\caption{Gamma-ray production differential cross section for $^{16}$O + $^{184}$W intereactions at $T_p=15$~A~MeV \cite{Breitbach1989}. The thick gray line is the hard photon fit using the parametrization in Eq.~(\ref{eq:BremsSpec}).\label{fig:BresSpec}}
\end{figure}

To conclude, Fig.~\ref{fig:BresSpec} shows a typical $\gamma$-ray spectrum that includes both the statistical and hard photons produced by $^{16}$O + $^{184}$W interactions at $T_p=15$~A~MeV \cite{Breitbach1989}. We see that the low energy statistical photons have larger cross sections compared to hard photons for $E_\gamma>30$~MeV. By increasing the collision energy the hard photon spectra becomes larger and harder. Hard photons dominate the $\gamma$-ray continuum for collision energies below $T_p<100$~A~MeV. Above this energy the $\pi^0$-meson production start to dominate.

\begin{table}[h]
\caption{References for the hard photon inverse slope parameter $E_0^\gamma$ experimental data. \label{tab:E0g}}
\begin{tabular}{lcr}
\toprule
System & $T_p$ [A~MeV] & Reference\\
\hline
C+Mo & 11 & \citet{Gossett1990}\\
O+W & 15 & \citet{Breitbach1989}\\
N+(C,Zn,Pb) & 20--40 & \citet{Stevenson1986}\\
Kr+Ni;Ta+Au & 29.5--60 & \citet{Martinez1995}\\
Pb+Au       & 29.5--60 & \citet{Martinez1995}\\
He+(C,Zn,Pb) & 25,53 & \citet{Tam1988}\\
Li+(Li,Pb);Ne+Mg & 30 & \citet{Tam1989}\\
Ar+(Ca,Pb) & 30 & \citet{Tam1989}\\
Ar+Au & 30 & \citet{Njock1986}\\
Ar+Gd & 44 & \citet{Hingmann1987}\\
Kr+(C,Ag) & 44 & \citet{Bertholet1987}\\
(Pb,Ta)+Au;Kr+Ni & 30--60 & \citet{Schutz(1997)}\\
C+C & 48--84 & \citet{Grosse1986a}\\
D+(C,Zn,Pb) & 53 & \citet{Tam1988}\\
Kr+Ni & 60 & \citet{Martinez1994}\\
Ar+(C,Al,Cu) & 85 & \citet{Njock1988}\\
Xe+Sn & 89,124 & \citet{Clayton1989}\\
Ar+(C,Au) & 95 & \citet{Schubert1994}\\
p+(C,O,Al,Cu,Pb) & 140 & \citet{Edgington1966}\\
Ar+Ca & 180 & \citet{Martinez1999}\\
\toprule
\end{tabular}
\end{table}

\section{Subthreshold pion production}
\subsection{Total production cross section $\sigma_\pi$ \label{sec:XSpi}}
The pion production cross section from nuclear collisions is well studied for a wide range of collision energies and for a variety of colliding nuclei. If pions in nucleus--nucleus ($A+B$) interactions are indeed produced through individual in-medium nucleon--nucleon ($N+N$) collisions, then their production cross sections or multiplicities should scale with the number of participating nucleons. Unlike the free nucleon--nucleon collisions the in-medium ones are enhanced by the Fermi motion of nucleons in the interactions zone. A common parametrization of the meson production cross section in $A+B$ collisions is \citep[see e.g.][]{Metag(1993)pi0_gamma}:
\begin{equation}\label{eq:XSpi}
\sigma_\pi=\sigma_R\,\left\langle A_{part}\right\rangle_b\,P_\pi,
\end{equation}
\noindent where $\sigma_R$ is the $A+B$ reaction cross section given in Eq.~(\ref{eq:XSr}), $P_\pi$ is the in-medium pion production probability per participant and $\left\langle A_{part}\right\rangle_b$ is the number of participants calculated within the geometrical model and averaged over the impact parameter $b$ \citep[see e.g.][]{Hufner1977,Cugnon1981}:
\begin{equation}
\left\langle A_{part} \right\rangle_b= \frac{A_p\,A_t^{2/3}+A_t\,A_p^{2/3}}{\left(A_p^{1/3}+A_t^{1/3}\right)^2}.
\end{equation}

\begin{figure}
\includegraphics[scale=0.47]{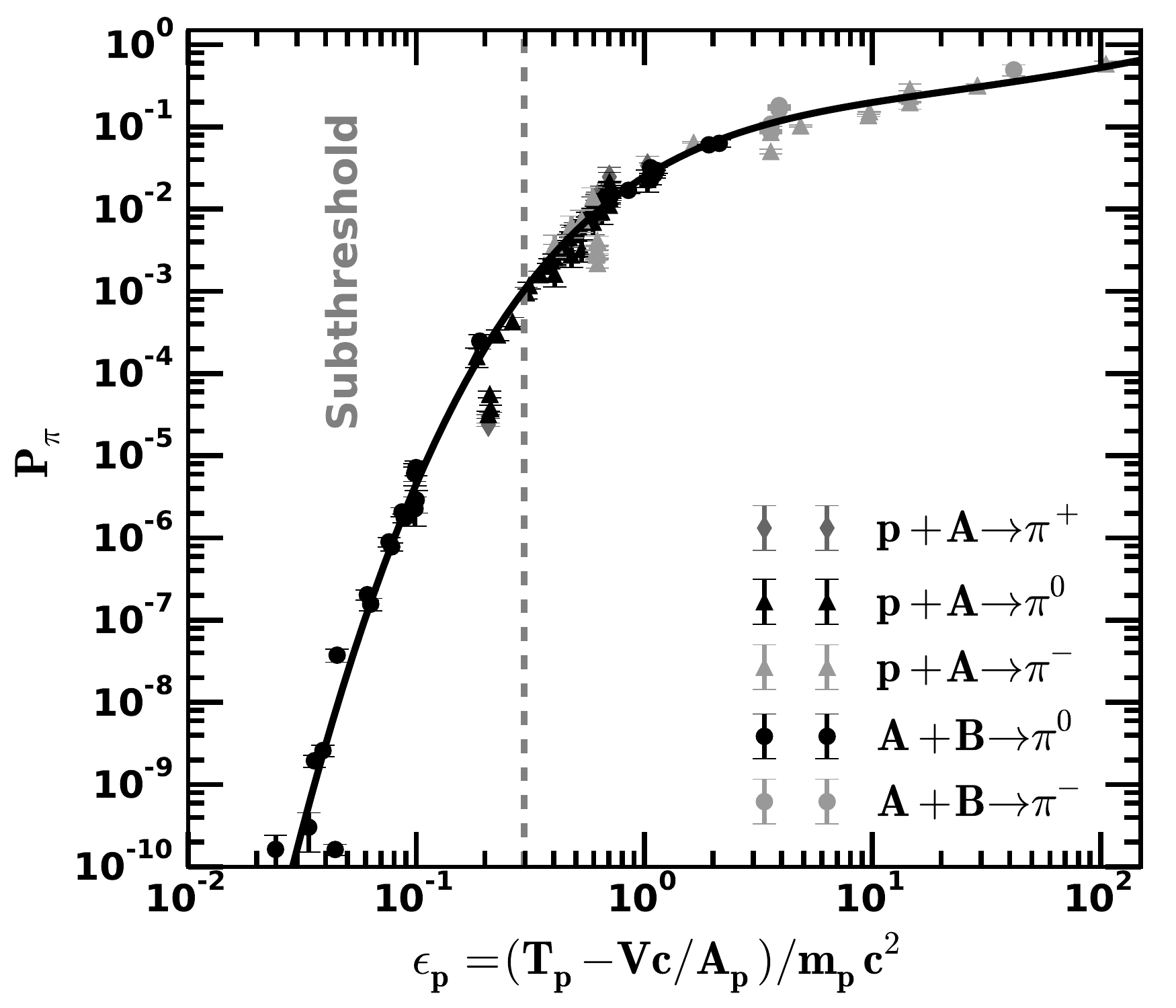}
\caption{The universal pion production probability as a function of $\epsilon_p$. All experimental data shown here are listed in Table~\ref{tab:XSpApi0}--\ref{tab:XSpABpimp}. The fitting curve is the parametrization given in Eq.~(\ref{eq:PpiFit}), $P_\pi/(\mathcal{I}_\pi \times \zeta)=\epsilon_p^{-1/4}\times \exp\left[\epsilon_p^{-1/4}\left( 0.0057\,x^4+ 0.019\,x^3 -0.19\,x^2 + 1.07\,x -3.7\right)\right]$ where $x=\log(\epsilon_p)$. \label{fig:Ppi}}
\end{figure}

I have collected public available experimental pion production cross section data that cover the nuclear collision energy range $20~{\rm A~MeV}<T_p<100~{\rm A~GeV}$ and for nuclei that are lighter than Zr. These data are recorded in Table~\ref{tab:XSpApi0}--\ref{tab:XSpABpimp}. Using Eq.~(\ref{eq:XSpi}) and these cross section data, one can compute the probability $P_\pi$, which is plotted in Fig.~\ref{fig:Ppi} as a function of $\epsilon_p=(T_p-V_c/A_p)/m_p$.

It is clear from Fig.~\ref{fig:Ppi} that the experimental $P_\pi$ data show a systematic increase with the collision energy per nucleon. I have parametrized the probability $P_\pi$ for $T_p\leq 100$~A~GeV as follows: 
\begin{equation}\label{eq:PpiFit}
\begin{split}
\Pi_4(x) &= a_0 + a_1\,x + a_2\,x^2 + a_3\,x^3 + a_4\,x^4\\
P_\pi&=\epsilon_p^{-1/4}\times\exp\left[\epsilon_p^{-1/4}\times \Pi_4(x)\right] \times \mathcal{I}_\pi \times \zeta,
\end{split}
\end{equation} 
\noindent where $x=\log(\epsilon_p)$. After fitting the experimental data one finds that $a_0=-(3.70\pm 0.05)$, $a_1= 1.07 \pm 0.06$, $a_2=-(0.19\pm 0.03)$, $a_3=(19\pm 7)\times10^{-3}$ and $a_4=(5.7\pm 1.8)\times10^{-3}$.

Different nuclear interactions produce different $\pi^+$, $\pi^0$ and $\pi^-$ yields. The function $\mathcal{I}_\pi$ takes into account these differences for a particular $A+B$ interaction by assuming that the differences arise due to isospin symmetry. Function $\mathcal{I}_\pi$ normalizes the probability $P_\pi$ for one of the $\pi$-mesons with respect to the $\pi^0$ one. Therefore, by adopting the isospin relations that exist between the three different pion yields from nucleon--nucleon interactions \citep[see e.g.][]{Golokhvastov2001a}, one finds:
\begin{equation}\label{eq:Isospin}
\mathcal{I}_\pi(\xi_p, \xi_t) = \begin{cases}
     (3 + \xi_p + \xi_t)/4 &  \text{for } \pi^+\\
     1 &  \text{for }\pi^0 \\
     (5 - \xi_p - \xi_t)/4 & \text{for } \pi^-\\
  \end{cases}
\end{equation}

\noindent where $\xi_p=Z_p/A_p$ and $\xi_t=Z_t/A_t$ give the ratio between the number of protons and the total number of nucleons for the projectile and target nuclei, respectively. Function $\zeta(T_p,A_p,A_t)$, on the other hand, ensures that pion production cross section for a given nucleus--nucleus interaction approaches zero when the collision energy approaches the absolute kinematic threshold, the kinetic energy per nucleon of which is given by:
\begin{equation}\label{eq:Tpth}
T_p^{\rm th} =\left(\frac{1}{A_p}+\frac{1}{A_t}\right)m_\pi + \frac{m_\pi^2}{2m_p\,A_pA_t},
\end{equation}
\noindent where $m_\pi$ is the pion mass. Based on the available experimental data near the kinematic threshold, a reasonably good approximation of $\zeta$ function is:
\begin{equation}
\zeta=\tanh\left( \max\left[0\,;\; 1-\left(\frac{T_p^{\rm th}}{T_p}\right)^3\right] \right)^{1/4},
\end{equation}

\noindent The function $\max$ makes $\zeta$ and $P_\pi$ equal zero for $T_p\leq T_p^{\rm th}$. The function $\zeta$ does not effect the shape of $P_\pi$ except near the pion absolute kinematic threshold.

The universal $P_\pi$ function that is plotted in Fig.~\ref{fig:Ppi} is actually the $P_\pi/(\mathcal{I}_\pi \times \zeta)$ as was defined in Eq.~(\ref{eq:PpiFit}).

\subsection{Differential cross section $d\sigma_\pi/dE_\pi$}
\begin{figure}
\includegraphics[scale=0.47]{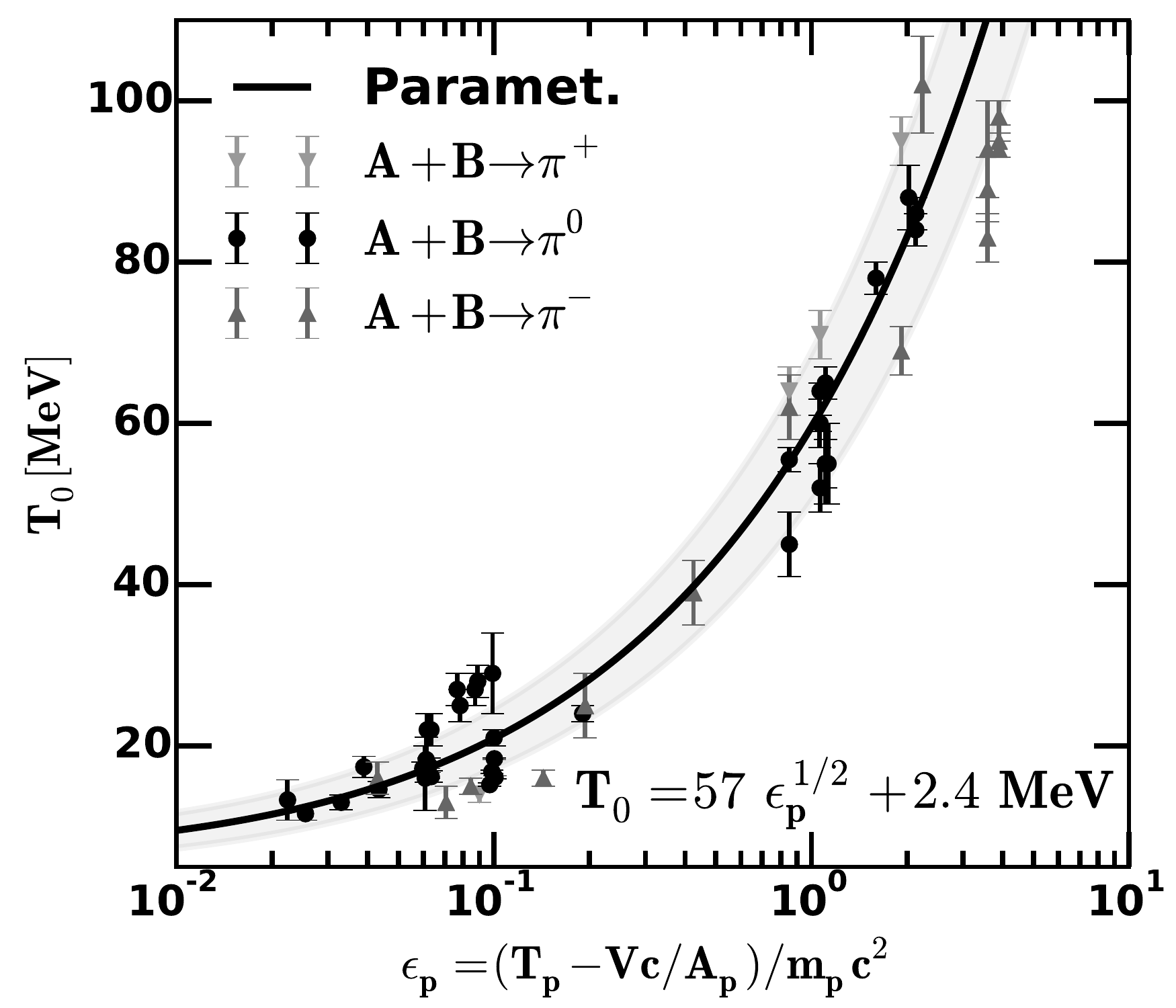}
\caption{The pion temperature as a function of $\epsilon_p$. The experimenal data points are listed in Table~\ref{tab:T0}. The line represents the parametrization given in Eq.~(\ref{eq:T0}) and the shaded area is the $1\sigma$ fitting bounds. \label{fig:T0}}
\end{figure}

Experimental observations for $T_p \lesssim 1$~A~GeV nuclear collisions show that the pion energy distribution has an exponential shape at high energies and it peaks at low energies at several tens of MeV. For head-on nucleus--nucleus collisions, the pion spectrum to a good approximation is isotropic; therefore, a statistical model is widely used to fit the experimental data \citep[see e.g.][]{Stock1986,Senger1999}. I assume here that the pion energy distribution for $T_p < 1$~A~GeV is given by a single relativistic Maxwell--J\"uttner distribution $f_{MB}(E_\pi,T_0)$, therefore, the pion differential cross section is:
\begin{equation}\label{eq:Pispec}
\frac{d\sigma_\pi}{dE_\pi}=\frac{\sigma_\pi}{m_\pi}\times f_{MB}(E_\pi,T_0).
\end{equation}
\noindent Where, $E_\pi$ is the pion total energy and $T_0$ is the pion temperature. Experiments show that the pion temperature increases systematically with the collision energy and they show that $T_0$ is almost independent on the pion species or the initial colliding nuclei. I have compiled here publicly available experimental pion temperature data for projectile energies below few GeV/nucleon and they are listed in Table~\ref{tab:T0}. For collision energies below 4~A~GeV, the following formula can parametrize reasonably well the pion temperature data:
\begin{equation}\label{eq:T0}
T_0=(57\pm8)\,\epsilon_p^{1/2}+(2.4\pm1.2)~{\rm MeV}.
\end{equation}
Figure~\ref{fig:T0} compares this parametrization with the available experimental data. 

Note that the $\Delta$-resonance start to be produced effectively at several hundreds of MeV per nucleon and it becomes non-negligible for collision energies around 1~A~GeV or higher. Its decays produce high energy pions that modify the tail of the Maxwellian distribution; therefore, Eq.~(\ref{eq:Pispec}) may not be a good representation of the high energy pions and their decaying products near the kinematic limit.

\begin{table}[h]
\caption{References to $p+A\to \pi^0$ experimental cross section data. \label{tab:XSpApi0}}
\begin{tabular}{lcr}
\toprule
System & $T_p$ [MeV] & Reference\\
\hline
p+(C,Al,Ni) & 201 & \citet{Bellini1989}\\
p+C & 470 & \citet{Bayukov1957}\\
p+D & 660 & \citet{Bayukov1957}\\
p+(Li,C,O,Al,Cu) & 665 & \citet{Dunaitsev(1964)}\\
p+He & 380--970 & \citet{Pollack1963}\\
\toprule
\end{tabular}
\end{table}

\begin{table}[h]
\caption{References to $A+B\to \pi^0$ experimental cross section data. \label{tab:XSABpi0}}
\begin{tabular}{lcr}
\toprule
System & $T_p$ [A~MeV] & Reference\\
\hline
(He,O)+Mg & 24--43 & \citet{Waters(1993)}\\
N+Al & 35 & \citet{BraunMunzinger(1984)}\\
O+Al & 38 & \citet{Julien1988}\\
Ar+Ca & 44 & \citet{Heckwolf(1984)}\\
C+(C,Ni) & 60--84 & \citet{Noll(1984)}\\
O+Al & 94 & \citet{Badala(1993)}\\
Ar+(Al,Ni) & 95 & \citet{Badala(1996)}\\
O+(Al,Ni) & 95 & \citet{Moisan1992}\\
Ar+Ca & 180 & \citet{Martinez1999}\\
Ar+Ca & 1000 & \citet{Schwalb1994}\\
(D,He,Ca)+Ca & 1040,1060 & \citet{Holzmann(1997)pi01GeV}\\
C+C & 1040,1060 & \citet{Holzmann(1997)pi01GeV}\\
C+C & 800--2000 & \citet{Averbeck2003}\\
C+C & 1000,1800 & \citet{Laue2000}\\
\toprule
\end{tabular}
\end{table}

\begin{table}[h]
\caption{References to charged pion production experimental cross section and multiplicity data. \label{tab:XSpABpimp}}
\begin{tabular}{lcr}
\toprule
System & $T_p$ [A~MeV] & Reference\\
\toprule
$\pi^+$ &  & \\
\hline
p+Ni & 201 & \citet{Badala1992}\\
p+He & 380--970 & \citet{Pollack1963}\\
p+(D,C,O,\\Al,Ni,Cu) & 585 & \citet{Crawford(1980)}\\
p+(Be,Al) & 13.7 $\times10^{3}$ &  \citet{Abbott1992}\\
p+Mg & 99.1 $\times10^{3}$ &  \citet{Whitmore1994}\\
\toprule
$\pi^-$ & & \\
\hline
p+He & 380--970 & \citet{Pollack1963}\\
p+(D,Be,C,O) & 585 & \citet{Crawford(1980)}\\
p+(Al,Ni,Cu) & 585 & \citet{Crawford(1980)}\\
p+C & 3.4$\times10^{3}$ & \citet{Agakishiyev(1985)pim}\\
C+(C,Ne,Si) & 3.7$\times10^{3}$ & \citet{Aksinenko1980}\\
C+(Cu,Zr) & 3.7$\times10^{3}$ & \citet{Aksinenko1980}\\
p+C & 9.1$\times10^{3}$ & \citet{Baatar1980}\\
p+C & 9.1$\times10^{3}$ & \citet{Armutliisky1987}\\
p+(Be,Al) & 13.7 $\times10^{3}$ &  \citet{Abbott1992}\\
p+Ne & 27.1 $\times10^{3}$ &  \citet{Miller1975}\\
p+Mg & 99.1 $\times10^{3}$ &  \citet{Whitmore1994}\\
p+Mg & 199.1 $\times10^{3}$ &  \citet{Brick1989}\\
(D,He,C)+C & 3.37$\times10^{3}$ & \citet{Agakishiyev(1985)pim}\\
\toprule
\end{tabular}
\end{table}

\begin{table}[h]
\caption{References to the experimental $\pi$-mesons temperature data. \label{tab:T0}}
\begin{tabular}{lcr}
\toprule
System & $T_p$ [A~MeV] & Reference\\
\toprule
$\pi^+$ &  & \\
\hline
C+C & 85 & \citet{Johansson1982}\\
Ni+Ni & 800--1800 & \citet{Muntz1997}\\
\toprule
$\pi^0$ & & \\
\hline
O+(Al,Ni) & 25 & \citet{Young(1986)}\\
Ar+(C,Al,Ni,Ag,Au),\\Au+Au & 25--95 & \citet{Piasecki(2010)pi0}\\
Xe+Au & 44 & \citet{Mayer(1993)pi0}\\
Kr+Ni;Ta+Au & 60 & \citet{Schutz(1997)}\\
C+(C,Ni) & 60--84 & \citet{Noll(1984)}\\
O+Al & 94 & \citet{Badala(1993)}\\
O+Al & 95 & \citet{Moisan1992}\\
Ar+Ca & 180 & \citet{Martinez1999}\\
Ar+Ca & 800 & \citet{Marin(1997)pi0}\\
Ar+Ca;Kr+Zr;Au+Au & 1000 & \citet{Schwalb1994}\\
(Ca,Ar)+Ca & 800--2000 & \citet{Averbeck2003}\\
C+C;Ni+Ni & 800--2000 & \citet{Averbeck2003}\\
(D,He,Ca)+Ca;C+C & 1040--1060 & \citet{Holzmann(1997)pi01GeV}\\
\toprule
$\pi^-$ & & \\
\hline
C+N & 41--135 & \citet{Suzuki(1991)}\\
Ar+KCl & 1800 & \citet{Brockmann1984}\\
C+Al & 183--2100 & \citet{Nagamiya(1982)}\\
(D,He,C)+C & 3.37$\times10^{3}$ & \citet{Backovic1992}\\
He+(Li,C);C+Ca & 3.66$\times10^{3}$ & \citet{Chkhaidze1992}\\
\toprule
\end{tabular}
\end{table}

\section{Gamma-rays and $e^\pm$ spectra\label{sec:3}}

Employing the kinematics of the pion decay and the pion differential cross section defined in Eq.~(\ref{eq:Pispec}) one can now calculate the $\gamma$-ray and $e^\pm$ pair production spectra for $T_p<1$~A~GeV. 

At higher energies, nuclear matter effects are expected to weaken; therefore, the pion production spectrum from nucleus--nucleus and nucleon--nucleon collisions are expected to be similar in shape. Recent high energy experiments find that the pion spectrum produced by $p+A$ and $A+B$ interactions in the forward hemisphere are similar to $pp$. Deviations from $pp$ are observed in the backward hemisphere where an excess of low energy pions are produced. A recent comprehensive study of the pion spectrum produced by $pp$ and $p+$C at 158~GeV/c show that the deviations between a two-component model -- that is constructed from the $pp$ data -- and the $p+$C pion data are less than 30--40~\% \citep{Barr2007}. In another study, a comparison of the experimental pion transverse mass distribution spectra below 1~GeV show that the deviations of Be+Be and Pb+Pb from $pp$ interactions with $T_p<100$~A~GeV is less than 20~\% and 40~\%, respectively \cite[see e.g.][]{Andronov2015CompareSpectra}. Combining these experimental findings, one may conclude that the expected deviations for the pion energy distribution shape between $pp$ and the light nuclear interactions that are relevant in astrophysics should not be larger than 30--40~\% for collisions with $T_p\leq100$~A~GeV. Thus, for simplicity I assume here that nucleus--nucleus and nucleon--nucleon collisions with $1\leq T_p\leq100$~A~GeV have identical pion spectral shape. As a result, the secondary particle production spectra for nucleus--nucleus collisions are calculated using their $pp$ spectra for which accurate parametrizations already exist \citep[see e.g.][]{Kamae2006,Kafexhiu2014}. 

\subsection{$\gamma$-ray differential cross section $d\sigma_\gamma/dE_\gamma$ \label{subsec:1}}
The main channels through which nucleus--nucleus interactions produce $\gamma$-rays with energy $E_\gamma>30$~MeV are the hard photons, $A+B\to \gamma$ and the neutral pion decay, $A+B\to\pi^0\to 2\gamma$. While the energy distribution of directly produced hard photons for a fixed projectile energy is described by Eq.~(\ref{eq:BremsSpec}), the distribution of $\pi^0\to2\gamma$ decay is determined by the $d\sigma_\pi/dE_\pi$ of the intermediate $\pi^0$-mesons given in Eq.~(\ref{eq:Pispec}) and by the kinematics of their decay:

\begin{equation}\label{eq:pi0decay}
\frac{d\sigma_\gamma}{dE_\gamma}=2\times \int\limits_{Y_\gamma}^{E_\pi^{\rm max}}\frac{d\sigma_\pi}{dE_\pi}\,\frac{dE_\pi}{P_\pi}.
\end{equation}
\noindent Here, the quantity $Y_\gamma$ is:
\begin{equation}
Y_\gamma=E_\gamma+\frac{m_\pi^2}{4\,E_\gamma}~,
\end{equation} 
\noindent and $E_\pi^{\rm max}$ is the maximum $\pi^0$ energy in the laboratory frame which is given by:
\begin{equation}\label{eq:EpiEgMAX}
\begin{split}
E_{\pi}^{\rm max} &=\gamma_{CM}\,(E_\pi^{CM} + P_\pi^{CM}\, \beta_{CM})\\
\gamma_{CM} &=\frac{A_p\,T_p + M_A+M_B}{\sqrt{s}}\\
s~~~ &= \left(M_A+M_B\right)^2 + 2\,M_B\left(A_p\,T_p+M_A\right)\\
E_\pi^{CM} &= \frac{s - \left(M_A+M_B\right)^2 + m_\pi^2}{2\sqrt{s}}.\\
\end{split}
\end{equation}

\noindent Where, $T_p$ is the projectile kinetic energy per nucleon, $s$ is the center of mass energy squared. $\beta_{CM}$, $\gamma_{CM}$ and $E_\pi^{CM}$ and $P_\pi^{CM}$ are the center of mass velocity, Lorentz factor and pion maximum energy and momentum, respectively. $M_A$ and $M_B$ are the mass of the projectile $A$ and the target $B$, respectively.

By performing the integration of Eq.~(\ref{eq:pi0decay}), one finds:
\begin{equation}
\begin{split}
\frac{d\sigma_\gamma}{dE_\gamma}= &\frac{2\,\theta_\pi\,\sigma_\pi}{m_\pi\,K_2(\theta_\pi^{-1})} \left[ \left(1+\frac{Y_\gamma}{T_0}\right) \times\exp\left(-\frac{Y_\gamma}{T_0}\right)-\right.\\
&\qquad\qquad \left. \left(1+\frac{E_\pi^{\rm max}}{T_0}\right)\times \exp\left(-\frac{E_\pi^{\rm max}}{T_0}\right) \right].
\end{split}
\end{equation}

\noindent The pion temperature $T_0$ is given in Eq.~(\ref{eq:T0}), $\theta_\pi=T_0/m_\pi$, $\sigma_\pi$ is the $\pi^0$ production cross section see Eq.~(\ref{eq:XSpi}), $Y_\gamma$ varies between $m_\pi\leq Y_\gamma \leq E_\pi^{\rm max}$ and $K_2(x)$ is the modified Bessel function of the second kind.

For collision energies $T_p\geq1$~A~GeV, the $\pi^0$ energy distribution for $A+B$ and nucleon--nucleon collisions are equal $f_{AB}^{\pi^0}=f_{NN}^{\pi^0}$. By averaging the pion spectrum over different nucleon--nucleon collisions (i.e. $pp$, $np$ and $nn$), it is shown in the Appendix~\ref{append:A} that $f_{AB}^{\pi^0}=f_{pp}^{\pi^0}$. Therefore, their respective $\pi^0\to2\gamma$ energy distributions should be equal $f_{AB}^{\gamma}=f_{pp}^{\gamma}$. To calculate $f_{pp}^{\gamma}$, the recent parametrization of the $pp\to\pi^0\to2\gamma$ production cross sections is adopted \cite{Kafexhiu2014}. Using the $\gamma$-ray production differential cross section $d\sigma^\gamma_{pp}/dE_\gamma$ and the pion production cross section $\sigma^\pi_{pp}$, the $\gamma$-ray energy distribution function is given by $f_{pp}^\gamma= (\sigma^\pi_{pp})^{-1}\times d\sigma^\gamma_{pp}/dE_\gamma$. As a result, the nucleus--nucleus $\gamma$-ray production differential cross section is given by $ d\sigma^\gamma_{AB}/dE_\gamma = \sigma^\pi_{AB}\times f_{pp}^\gamma$, where, $\sigma^\pi_{AB}$ is calculated using Eq.~(\ref{eq:XSpi}).

\subsection{$e^\pm$ pair production spectra\label{subsec:epm}}
After production, charged pions quickly decay into muons which are unstable and further decay into electrons and positrons, $\pi^\pm\to \mu^\pm \to e^\pm$. Although, electrons and positrons cannot be detected directly from an astrophysical source, they however, can emit $\gamma$-rays through radiative processes such as e.g. bremsstrahlung and synchrotron radiation. 

The decay kinematics of charged pions into $e^\pm$ is more complex than $\pi^0\to2\gamma$ because it involves the muon spin and the three body decay kinematics. A useful quantity that is found in the literature, is the $e^\pm$ energy distribution for a fixed pion energy \citep[see e.g.][]{Scanlon1965,Dermer1986}. This quantity is convolved with the pion spectrum to obtain the $e^\pm$ production spectra.

The charged pion energy distribution for $A+B$ collisions for $T_p<1$~A~GeV is described by Eq.~(\ref{eq:Pispec}). Assuming that the projectile $A$ flux is given by $J_A(T_p)$ and the target number density is $n_B$, we can calculate the pion production spectrum:
\begin{equation}\label{eq:dNdEpiepm}
\frac{dN_\pi}{dE_\pi}=4\pi\,n_B \int\limits_{T_p^{\rm th}}^\infty dT_p\;J_A(T_p)\; \frac{d\sigma_\pi^{AB}}{dE_\pi}(T_p,E_\pi).
\end{equation}

By convolving this quantity with the $e^\pm$ normalized energy distribution for a single pion energy $\Phi(\gamma_\pi,E_e)$ \cite{Dermer1986}, one can calculate the $e^\pm$ emissivity as follows:
\begin{equation}\label{eq:dNdEe}
\frac{dN}{dE_e}=\int\limits_{\bar{\gamma}_\pi}^\infty d\gamma_\pi \;\frac{dN_\pi}{d\gamma_\pi}(\gamma_\pi)\times \Phi(\gamma_\pi,E_e).
\end{equation}
\noindent Where, $\gamma_\pi=E_\pi/m_\pi$, $\bar{\gamma}_\pi = 1$ if $E_e<E_e^{\rm max}$ and $\bar{\gamma}_\pi = \frac{1}{2}(E_e/E_e^{\rm max}+E_e^{\rm max}/E_e)$ if $E_e>E_e^{\rm max}$. $E_e^{\rm max}=m_\mu \gamma_\mu(1+\beta_\mu)/2$, $\gamma_\mu=(m_\pi^2+m_\mu^2)/2 m_\pi m_\mu$, $\beta_\mu=(1-\gamma_\mu^{-2})^{1/2}$ and $m_\mu$ is the muon mass.

At higher collision energies ($T_p\geq 1$~A~GeV) the charged pion energy distribution from $A+B$ interactions averaged over different nucleon--nucleon collisions is $f_{AB}^{\pi^\pm}=\bar{\xi}\, f_{pp}^{\pi^\pm} + (1-\bar{\xi})\, f_{pp}^{\pi^\mp}$, see Appendix~\ref{append:A}. Therefore, their respective $e^\pm$ energy distributions that result from $\pi^\pm$ decay, should satisfy the same relations $f_{AB}^{e^\pm}=\bar{\xi}\, f_{pp}^{e^\pm} + (1-\bar{\xi})\, f_{pp}^{e^\mp}$. Here, $\bar{\xi}=(\xi_p +\xi_t)/2$ where $\xi_p=Z_p/A_p$ and $\xi_t=Z_t/A_t$ are the ratios of the number of protons to the total number of nucleons for the projectile and target nuclei, respectively. For calculating $f_{pp}^{e^\pm}$, the parametrization of the $pp\to e^\pm$ production cross sections \cite{Kamae2006} are adopted. The $e^\pm$ energy distributions are given by $f_{pp}^{e^\pm}=(\sigma^{e^\pm}_{pp})^{-1}\times d\sigma^{e^\pm}_{pp}/dE_e$ where $\sigma^{e^\pm}_{pp}=\int dE_e\,d\sigma^{e^\pm}_{pp}/dE_e$. The $A+B\to e^\pm$ differential cross section is $d\sigma^{e^\pm}_{AB}/dE_e=\sigma^{\pi^\pm}_{AB}\times \left[\bar{\xi}\, f_{pp}^{e^\pm} + (1-\bar{\xi})\, f_{pp}^{e^\mp} \right]$ where $\sigma^{\pi^\pm}_{AB}$ is the charged pion production cross section for a given $A+B$ interaction shown in Eq.~(\ref{eq:XSpi}).

\section{Comparison with the experimental data}
In this section, the parametrizations developed so far are compared with the available experimental data for nuclear interactions that are relevant in astrophysics.

\begin{figure}
\includegraphics[scale=0.5]{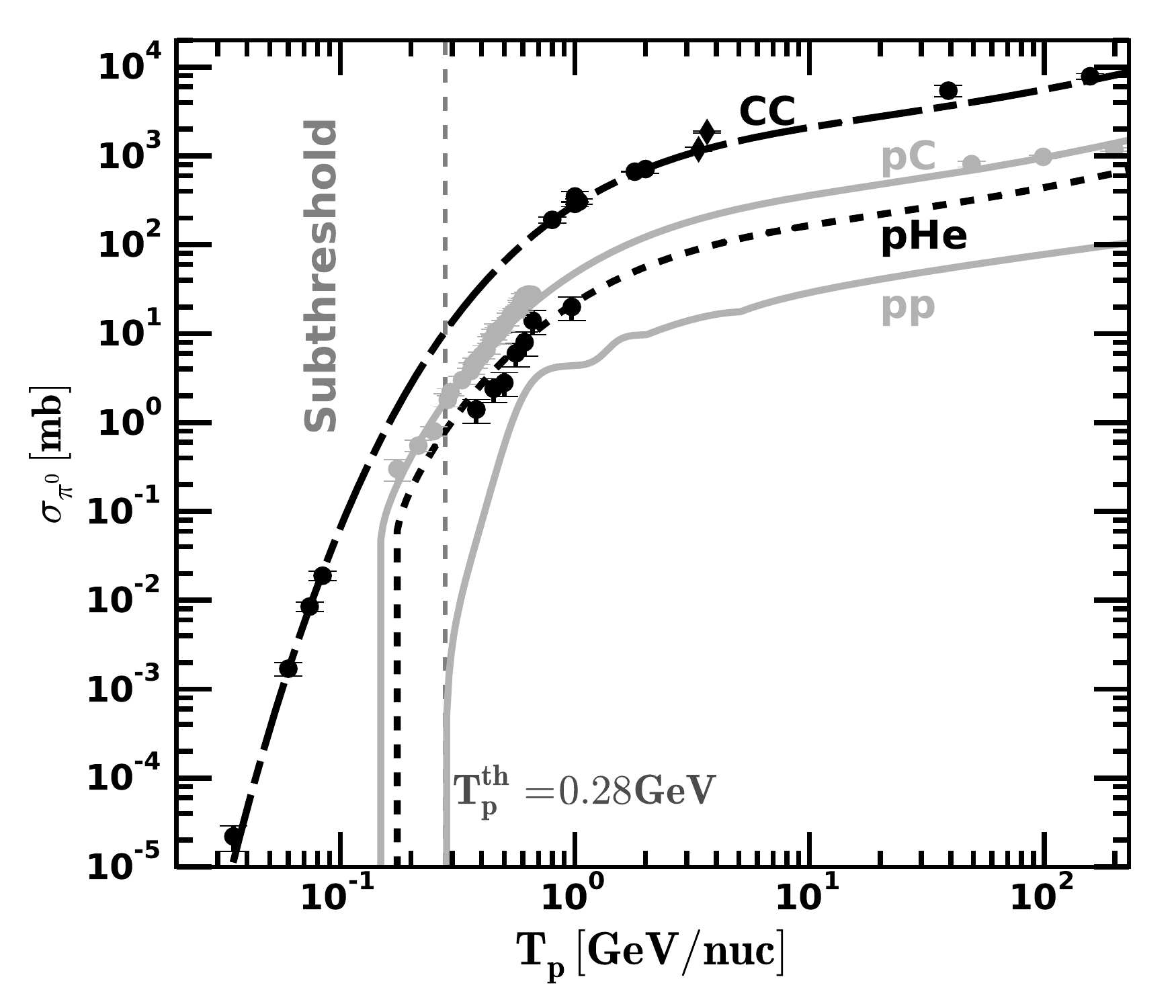}
\caption{Neutral pion production cross section as a function of the projectil kinetic energy per nucleon for $p+\rm{^4He}$, $p+\rm{^{12}C}$ and $\rm{^{12}C+^{12}C}$ interactions. The experimental data points are described in the text, whereas the curves are the predictions of Eq.~(\ref{eq:XSpi}). The $p+p\to\pi^0$ production cross section is added for comparison. \label{fig:XSpi0Valid}}
\end{figure}

Figure~\ref{fig:XSpi0Valid} compares the $\pi^0$ production cross section for $p+\rm{^4He}$, $p+\rm{^{12}C}$ and $\rm{^{12}C+^{12}C}$ interactions. For comparison the $pp\to\pi^0$ production cross section \cite{Kafexhiu2014} is plotted, too. The references for the experimental data points are found in Table~\ref{tab:XSpApi0}--\ref{tab:XSpABpimp}. The three high energy data points for $p+\rm{^{12}C}$ at $P_p=50$, 100 and 200~GeV/c are taken from \citep{Elias1980} and are not direct measurements of the $\pi^0$ production yield. These high energy data are obtained from the total charge particle yield which is dominated by $\pi^\pm$ yields. Thus, the average $\pi^0$ production multiplicity is calculated using the isospin symmetry $\left\langle \pi^0 \right\rangle = (\left\langle \pi^+ \right\rangle + \left\langle \pi^- \right\rangle)/2$. The experimental data for $\rm{^{12}C+^{12}C}$ at $P_p=40$ and 158~A~GeV/c are taken from \citep{NA49CC40,NA49CC158} and correspond to charged pions yields. Assuming that the isospin symmetry holds, then the yields for charged and neutral pions should be similar because both the projectile and target have equal number of protons and neutrons, see Appendix~\ref{append:A}. 

\begin{figure}
\includegraphics[scale=0.45]{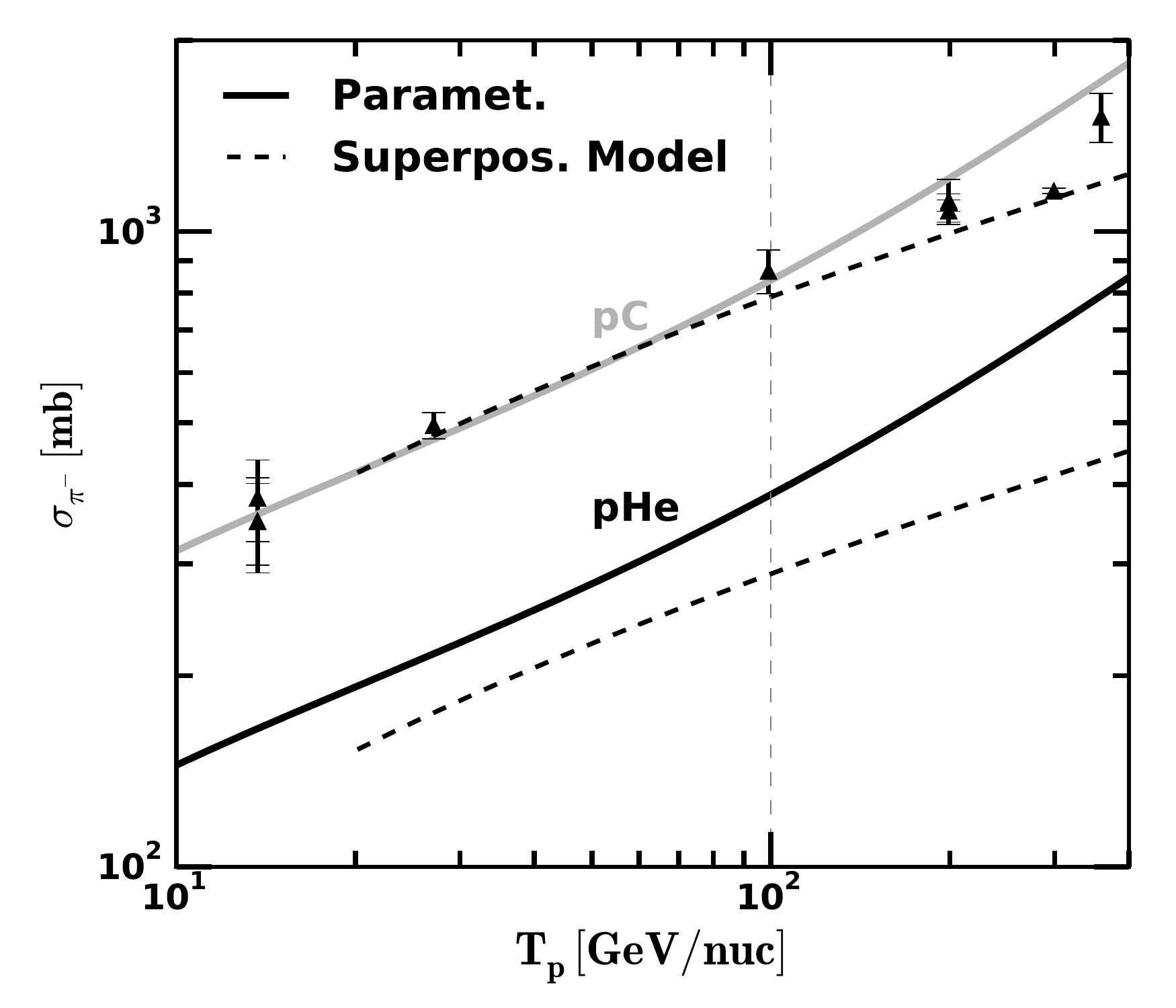}
\caption{Negativ pion production cross section as a function of projectil kinetic energy per nucleon for $p+\rm{^4He}$ and $p+\rm{^{12}C}$ interactions. The experimental data points are described in the text, the full lines are described by Eq.~(\ref{eq:XSpi}), whereas the dash lines represent the superposition model \citep{Elias1980} (see the text). \label{fig:XSpimValid}}
\end{figure}

Figure~\ref{fig:XSpi0Valid} shows that the parametrization formula presented in Eq.~(\ref{eq:XSpi}) fits very well the experimental data for $T_p \leq 100$~A~GeV. One can even extrapolate this parametrization to few-hundreds of A~GeV without causing large uncertainties, see Fig.~\ref{fig:XSpimValid}. 

Figure~\ref{fig:XSpimValid} compares the parametrization of the $\pi^-$ production cross section with the available $p+\rm{^{12}C}\to \pi^-$ production yields from different experiments. The superposition model curve is calculated using the pion production yield ratio $R_{pA}$. The average negative pion multiplicity for $p+C$ is calculated as $\left\langle \pi^- \right\rangle_{pC} = R_{pC}\,\left\langle \pi^- \right\rangle_{pp}$, where $\left\langle \pi^- \right\rangle_{pp}$ is taken from \cite{Golokhvastov2001a}. The ratio $R_{pA}$ is parametrized experimentally as $R_{pA}\approx 0.5 + 0.58\,\bar{\nu}$ \citep{Elias1980}. The $\bar{\nu}$ is the average number of inelastic interactions and is given by $\bar{\nu}=A\,\sigma_{pp}/\sigma_{pA}\approx 0.66\,A^{0.31}$, where $A$ is the target mass number and $\sigma_{pp}$ and $\sigma_{pA}$ are the absorption cross sections for nucleon--nucleon and nucleon--nucleus interactions, respectively \citep{Elias1980}. It is clear from Fig.~\ref{fig:XSpimValid} that the parametrization and the superposition model are in good agreement for $T_p \leq 100$~A~GeV. Their differences for $p+\rm{^4He}$ are less than 15~\%.

\begin{figure}
\includegraphics[scale=0.45]{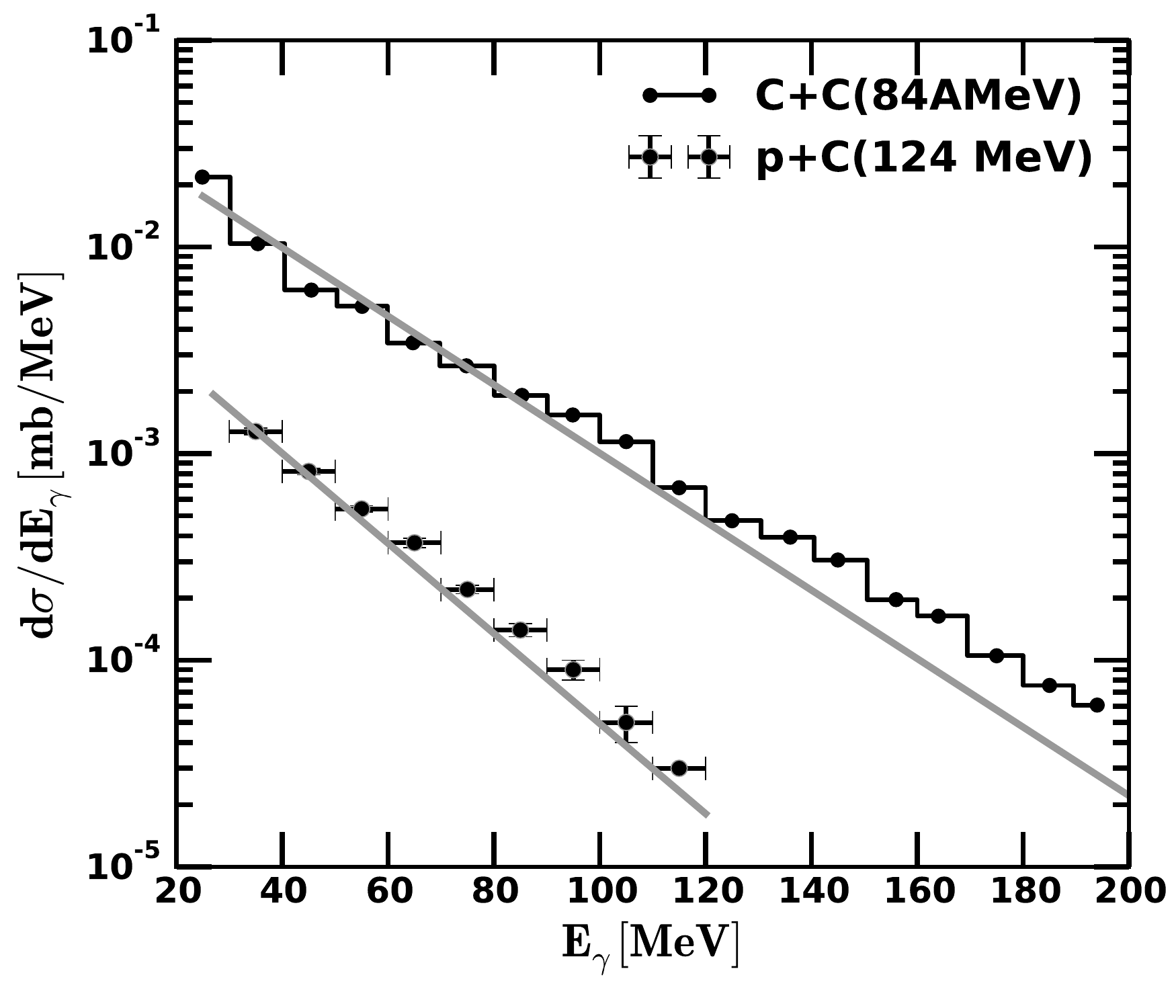}
\caption{Comparison of the hard photon $\gamma$-ray production differential cross section Eq.~(\ref{eq:BremsSpec}) with the available experimental data for $p+\rm{^{12}C}\to\gamma$ at 124~MeV \cite{Edgington1966} and $\rm{^{12}C+^{12}C}\to\gamma$ at 84~A~MeV \cite{Grosse1986a}. \label{fig:ValidBrems}}
\end{figure}

Figure~\ref{fig:ValidBrems} compares the hard photon parametrization formula given in Eq.~(\ref{eq:BremsSpec}) with the experimental data for $p+\rm{C}$ at $T_p=124$~MeV \cite{Edgington1966} and for $\rm{C+C}$ at $T_p=84$~A~MeV \cite{Grosse1986a}. As we can see, the parametrization described here fits reasonably well these experimental data. 

\begin{figure}
\includegraphics[scale=0.45]{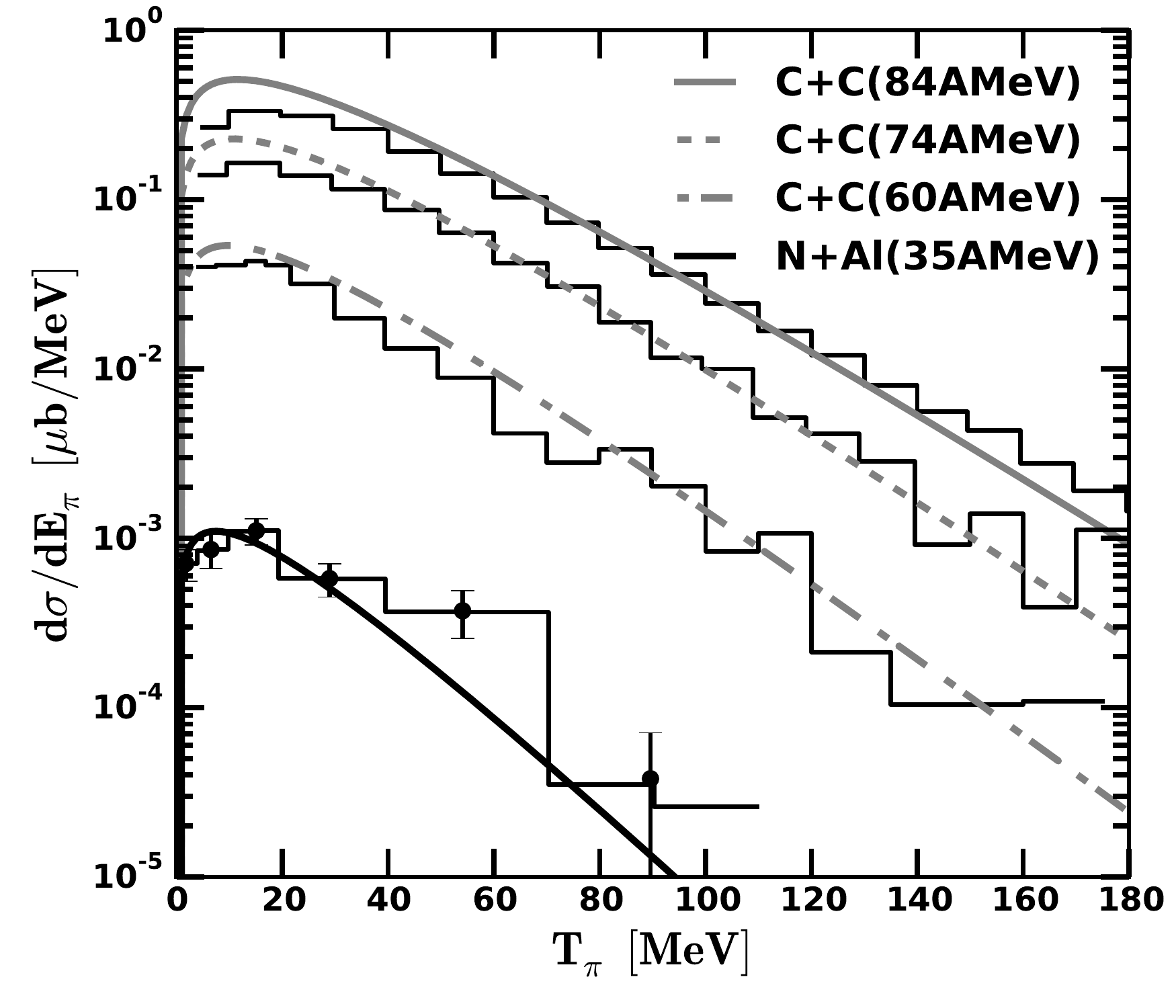}
\caption{Comparison of the subthreshold $\pi^0$ production differential cross section for $\rm{^{14}N+^{27}Al}\to\pi^0$ at 35~A~MeV \citep{BraunMunzinger(1984)} and $\rm{^{12}C+^{12}C}\to\pi^0$ for 60, 74 and 84~A~MeV \cite{Noll(1984)}. The histogram line represent the experimental data, whereas the full line is the parametrization presented in Eq.~(\ref{eq:Pispec}). \label{fig:Lowpi0}}
\end{figure}

Figure~\ref{fig:Lowpi0} compares the low energy pion differential cross section parametrization given in Eq.~(\ref{eq:Pispec}) with the available experimental data for $\rm{^{14}N+^{27}Al}\to\pi^0$ at 35~A~MeV \citep{BraunMunzinger(1984)} and $\rm{^{12}C+^{12}C}\to\pi^0$ at 60, 74 and 84~A~MeV \cite{Noll(1984)}. As we can see, the parametrization fits well the data at such energies. 

\begin{figure*}
\includegraphics[scale=0.33]{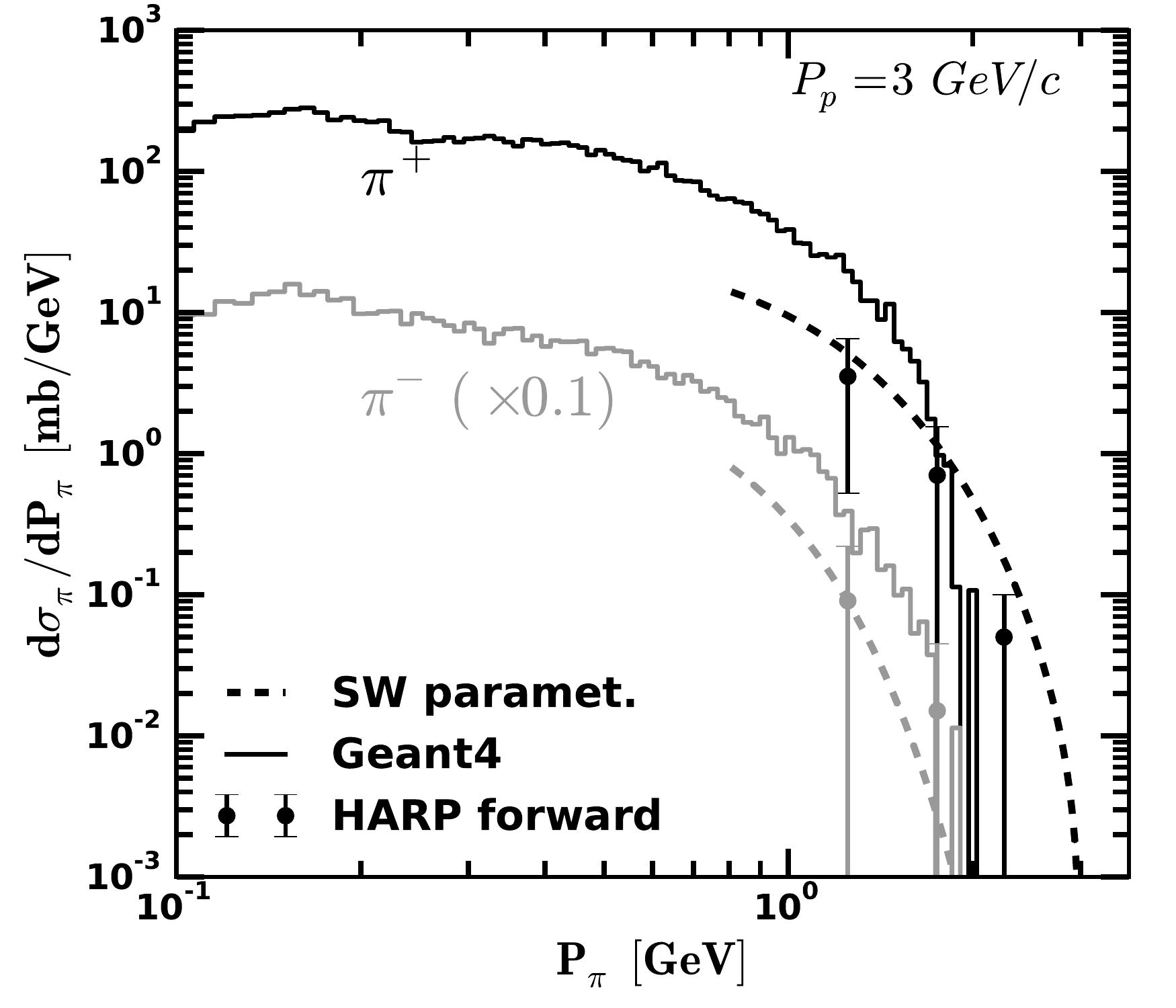}
\includegraphics[scale=0.33]{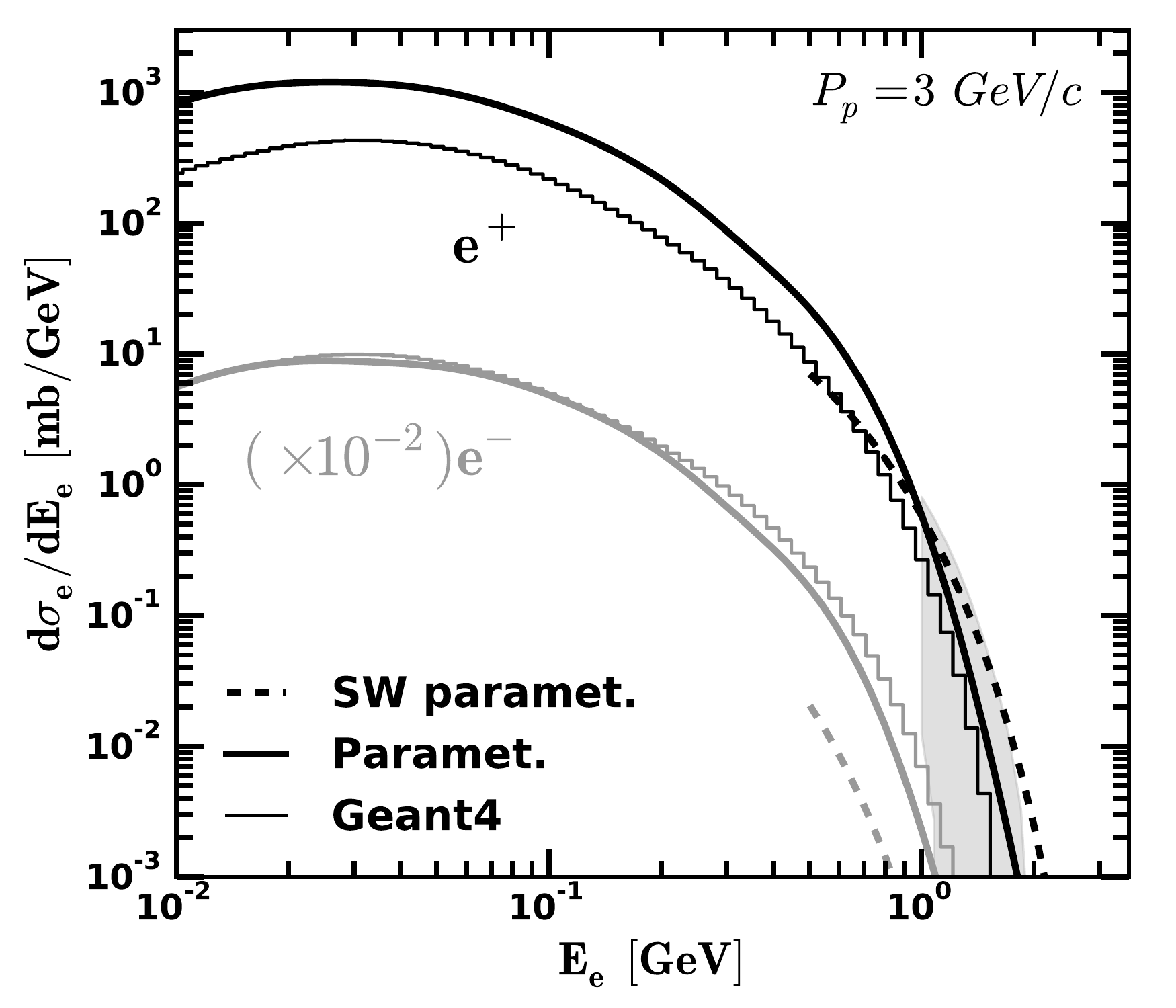}
\includegraphics[scale=0.33]{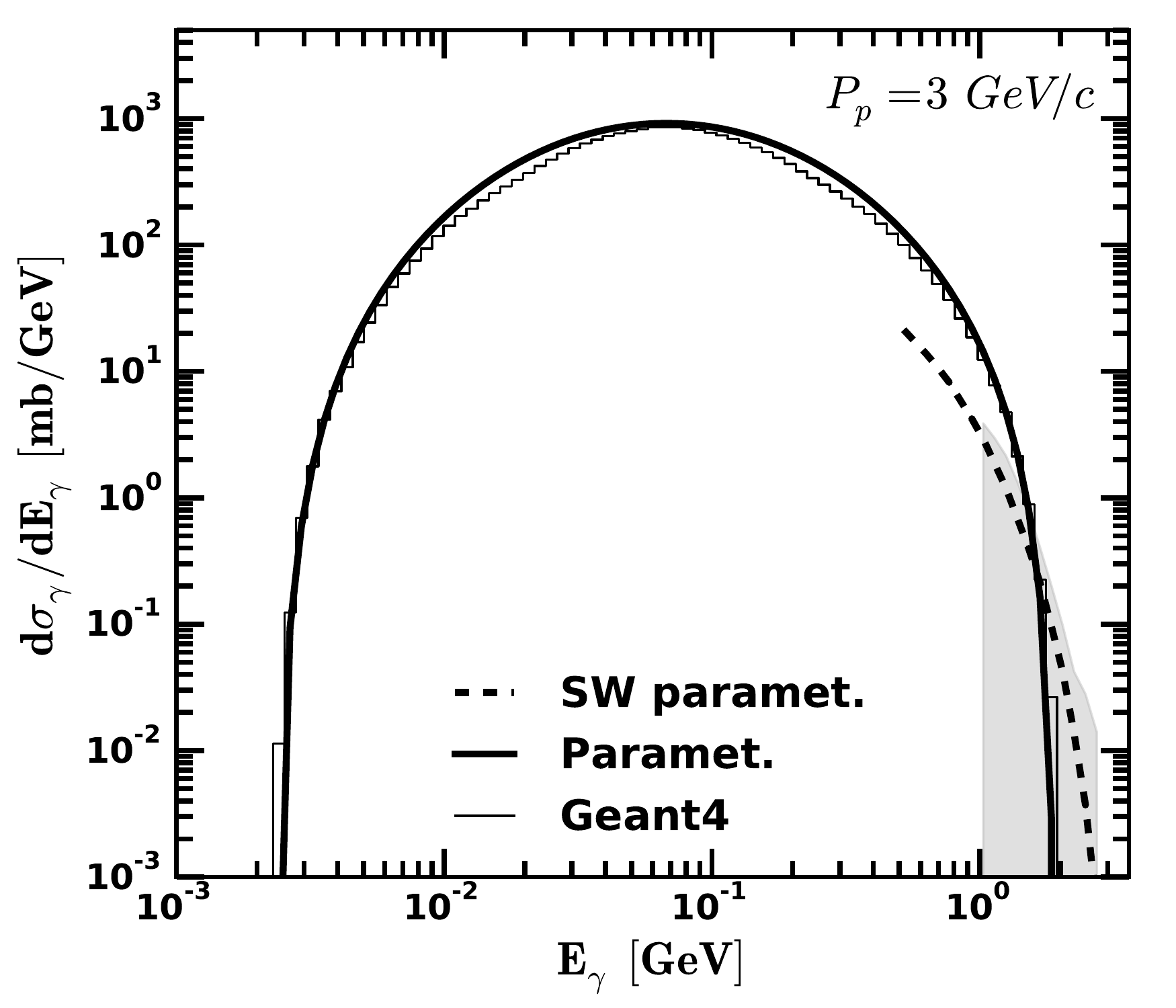}\\
\includegraphics[scale=0.33]{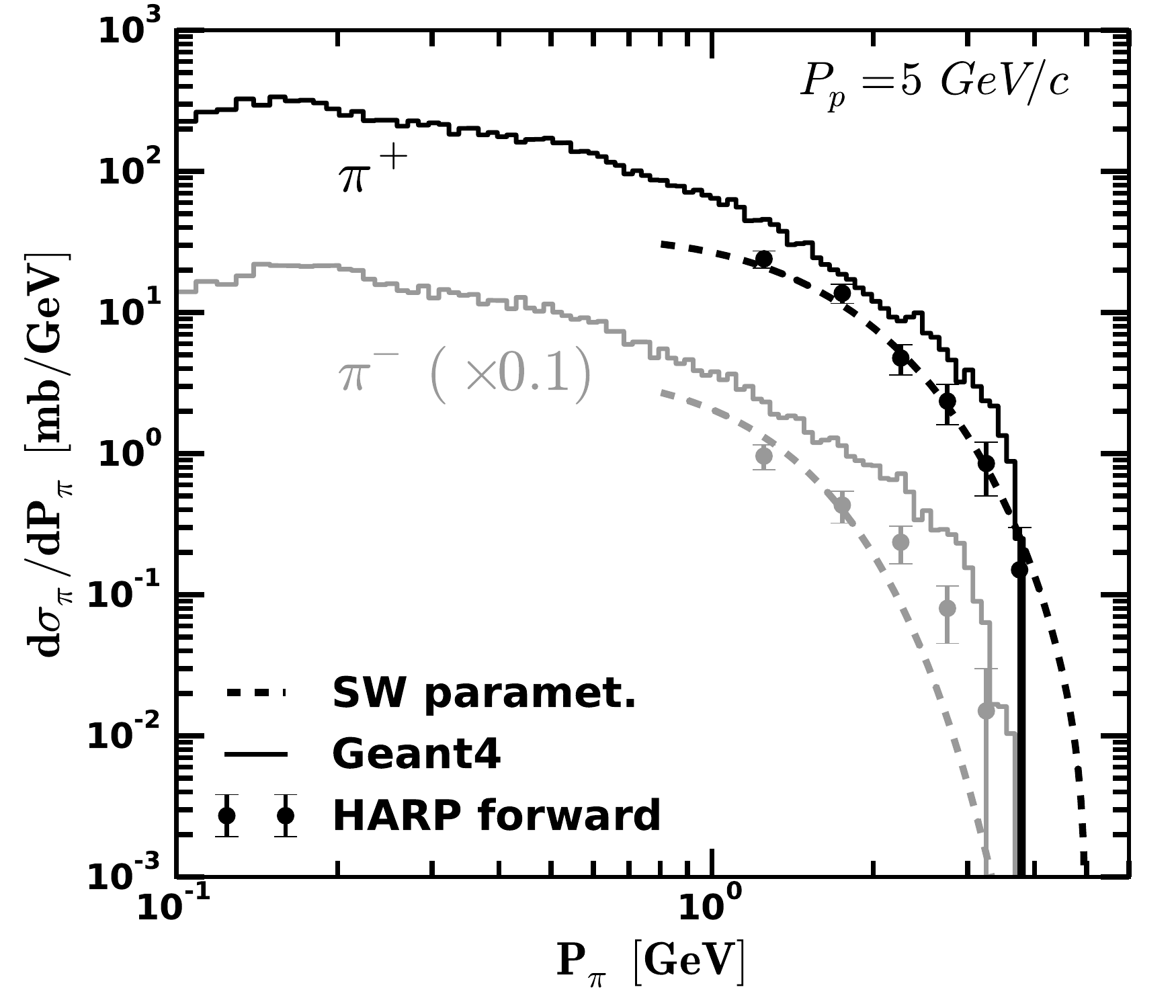}
\includegraphics[scale=0.33]{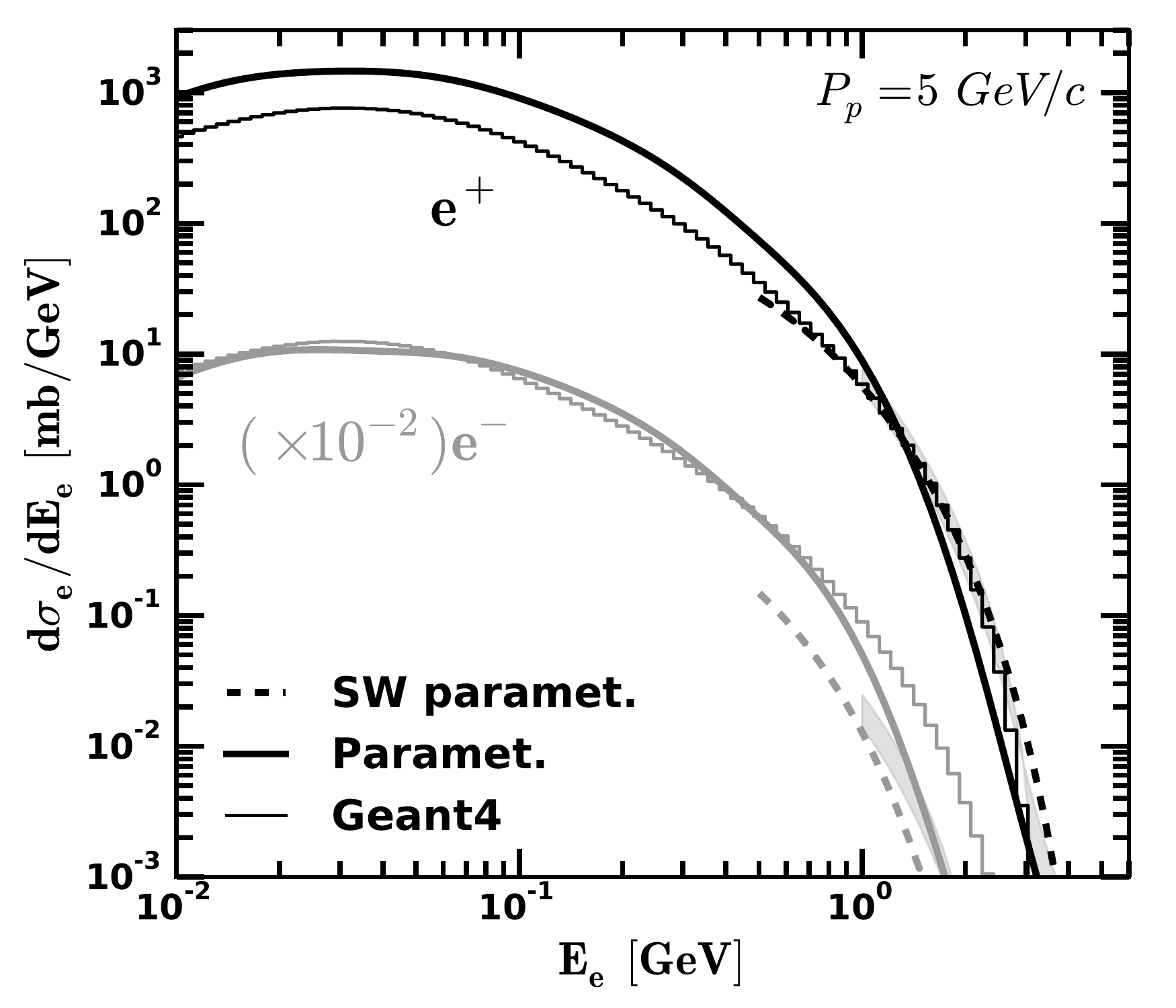}
\includegraphics[scale=0.33]{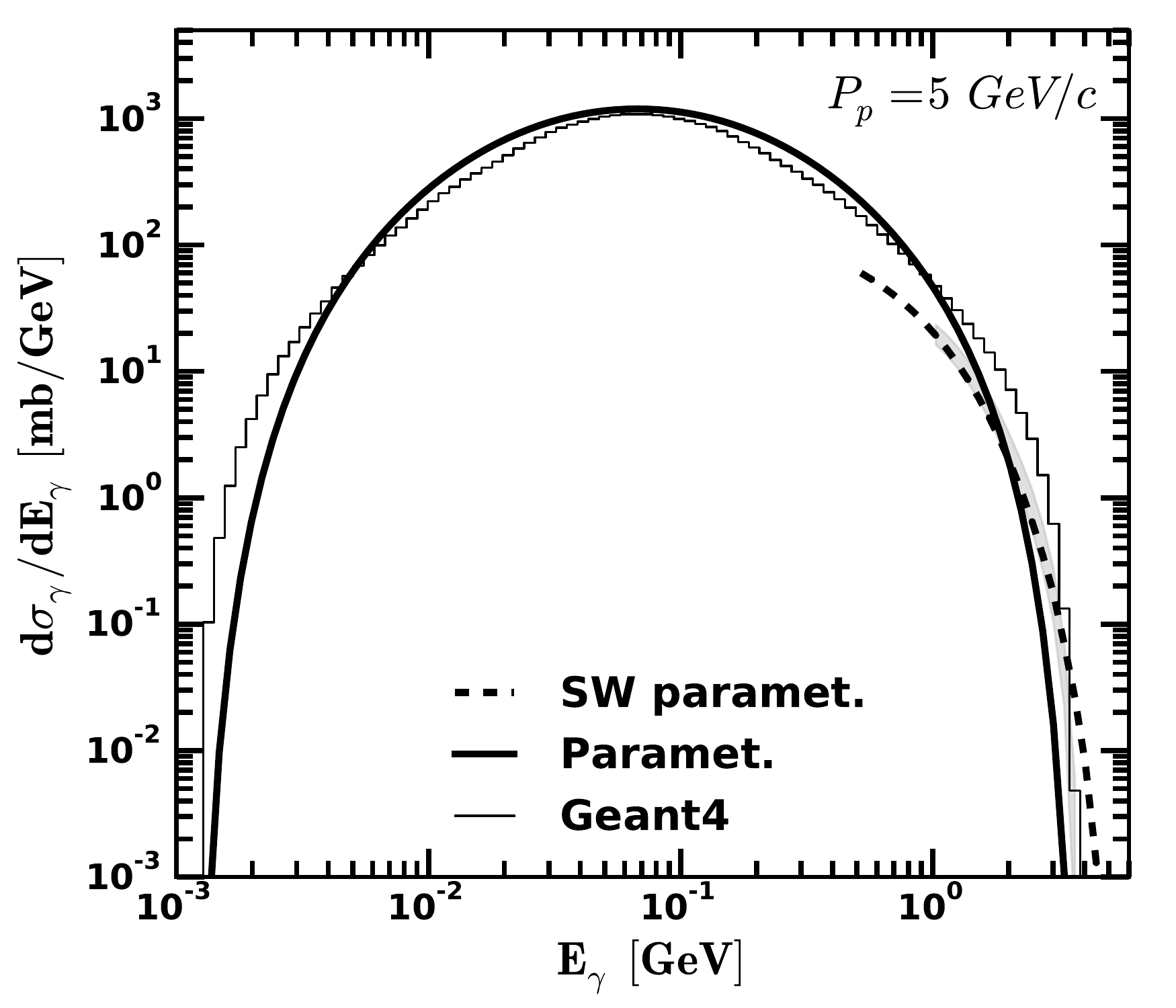}\\
\includegraphics[scale=0.33]{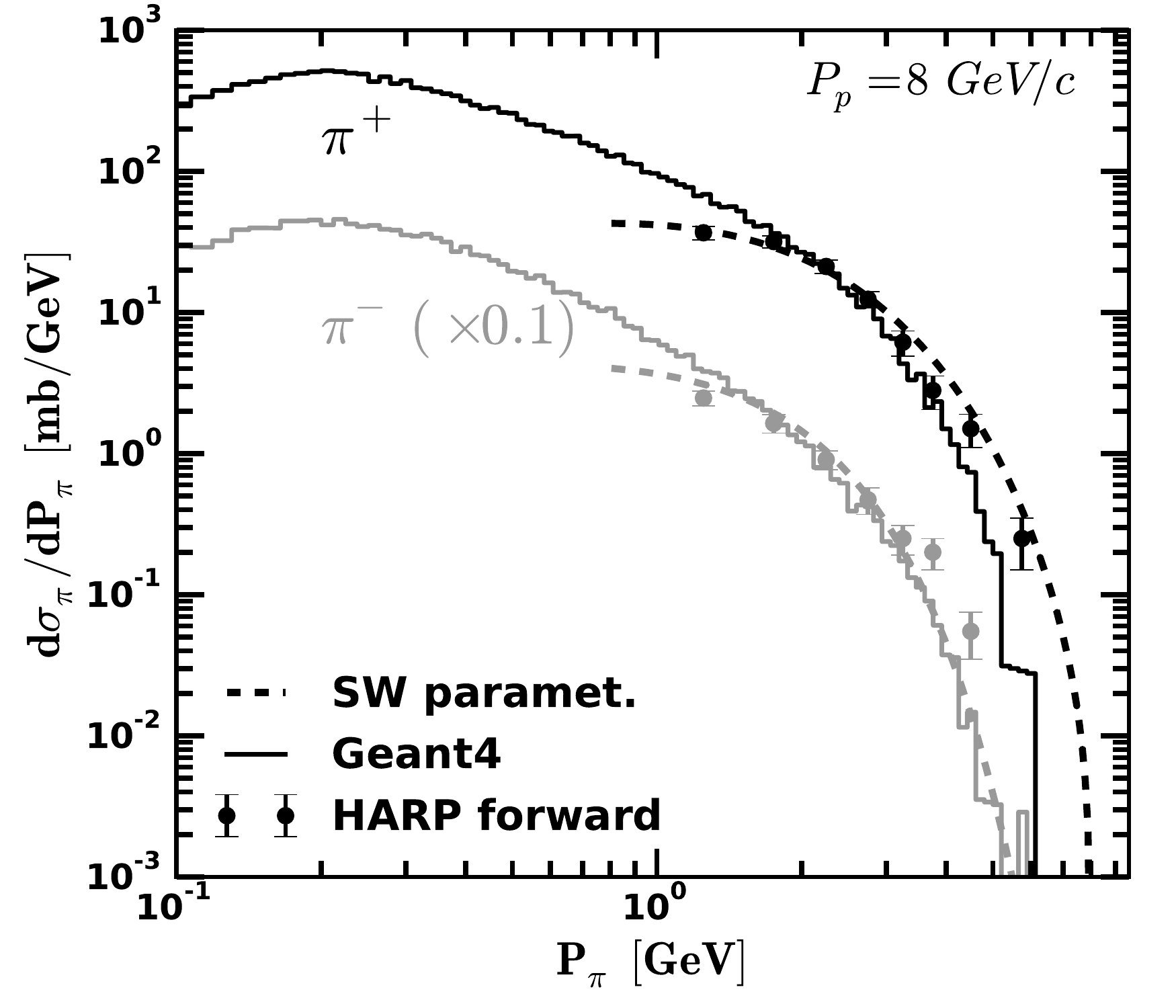}
\includegraphics[scale=0.33]{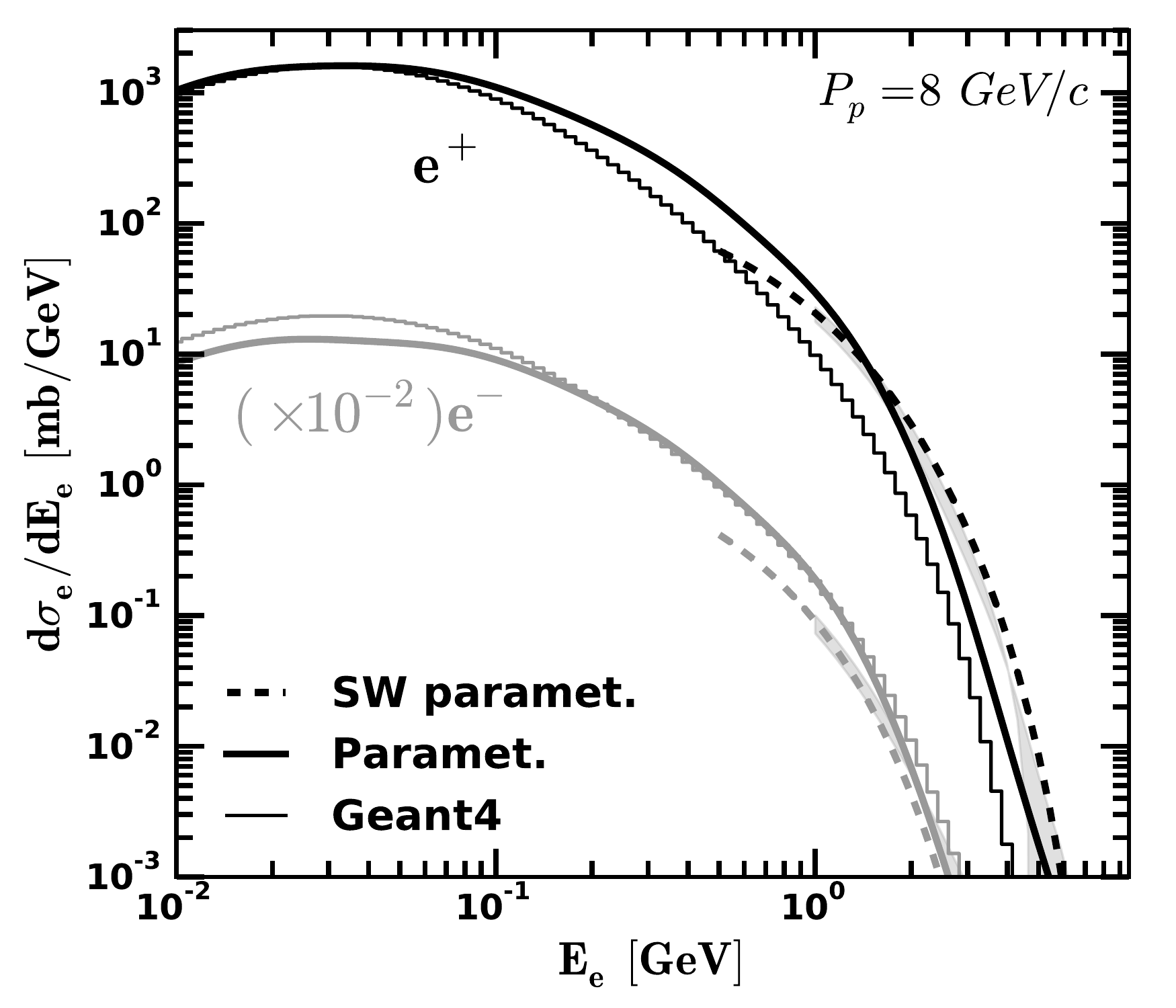}
\includegraphics[scale=0.33]{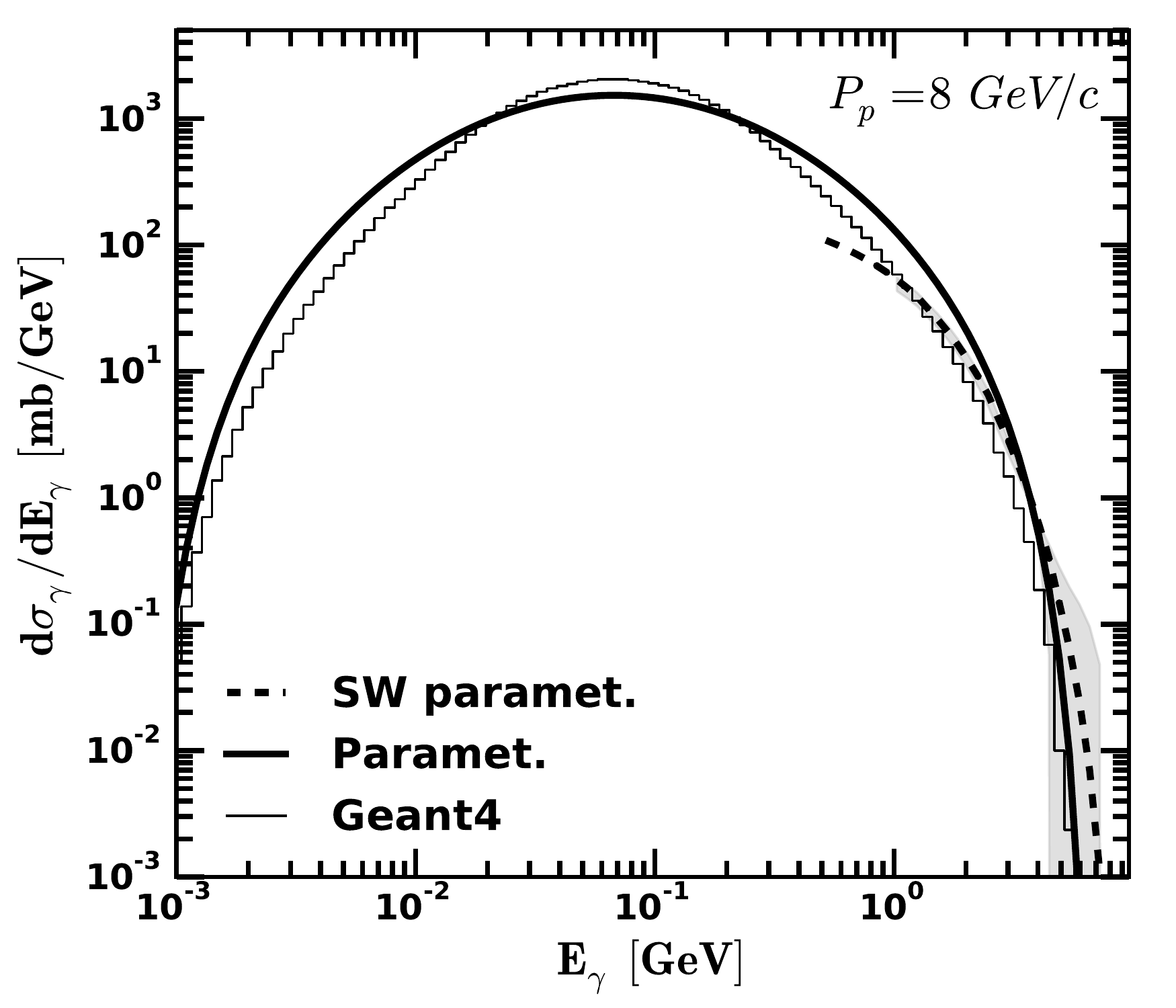}\\
\includegraphics[scale=0.33]{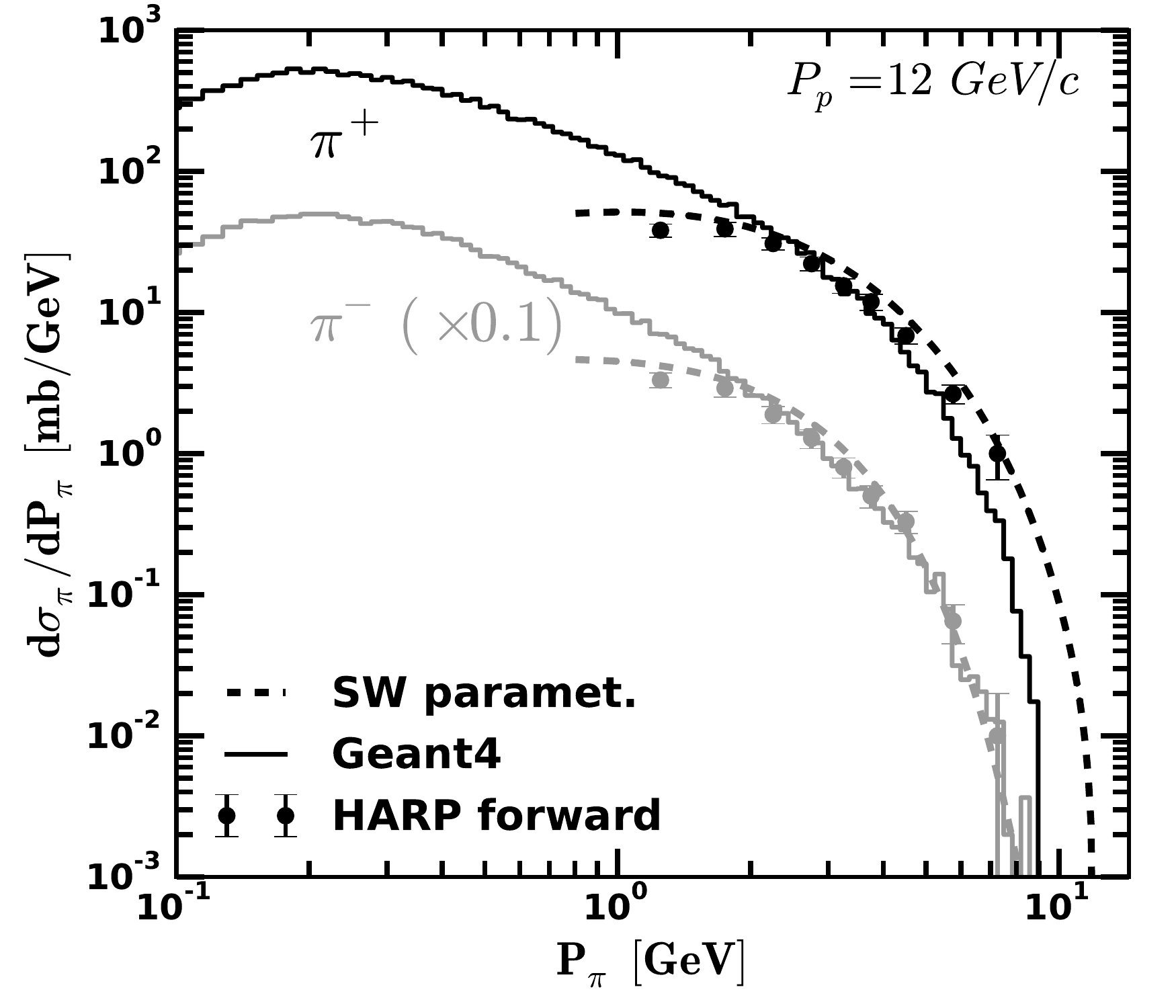}
\includegraphics[scale=0.33]{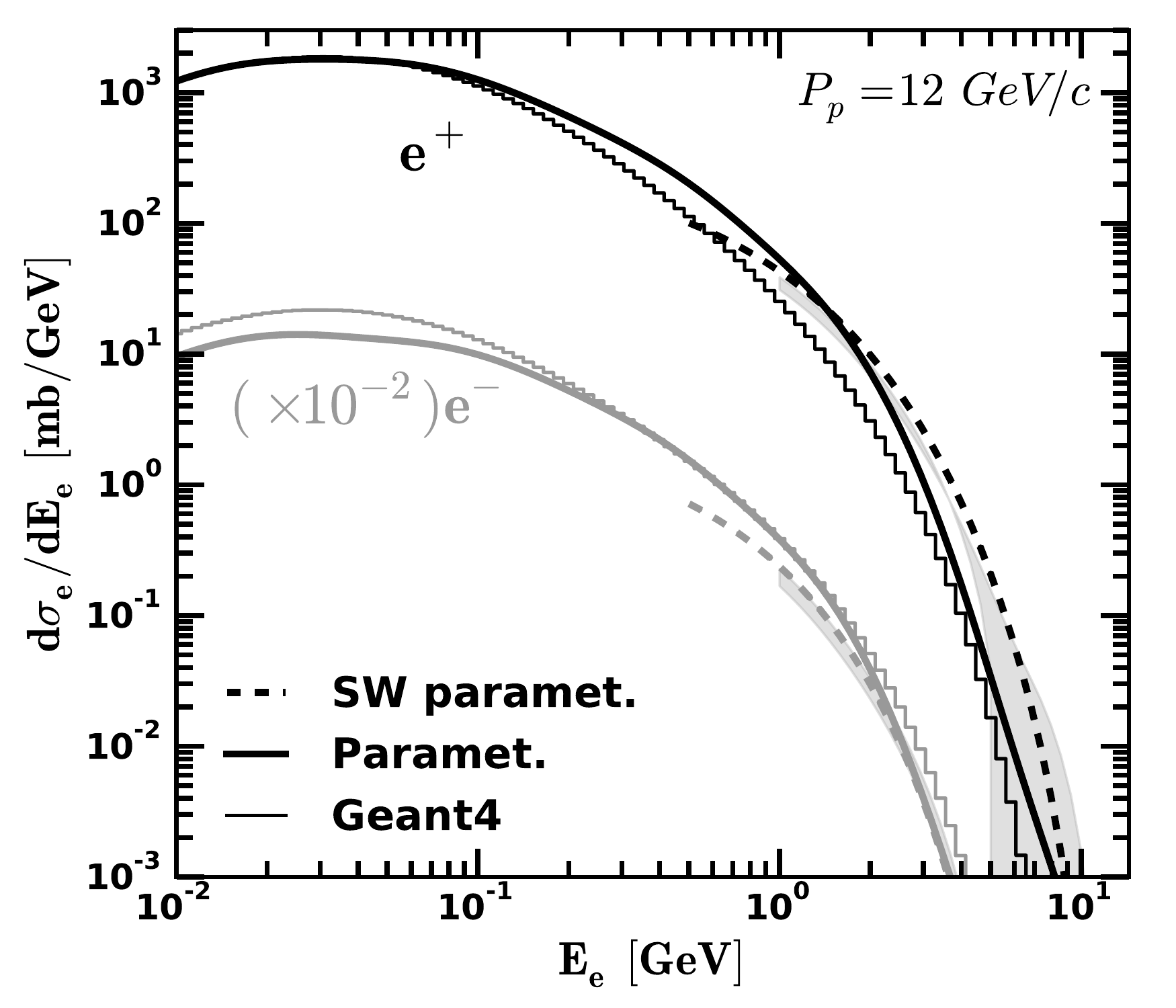}
\includegraphics[scale=0.33]{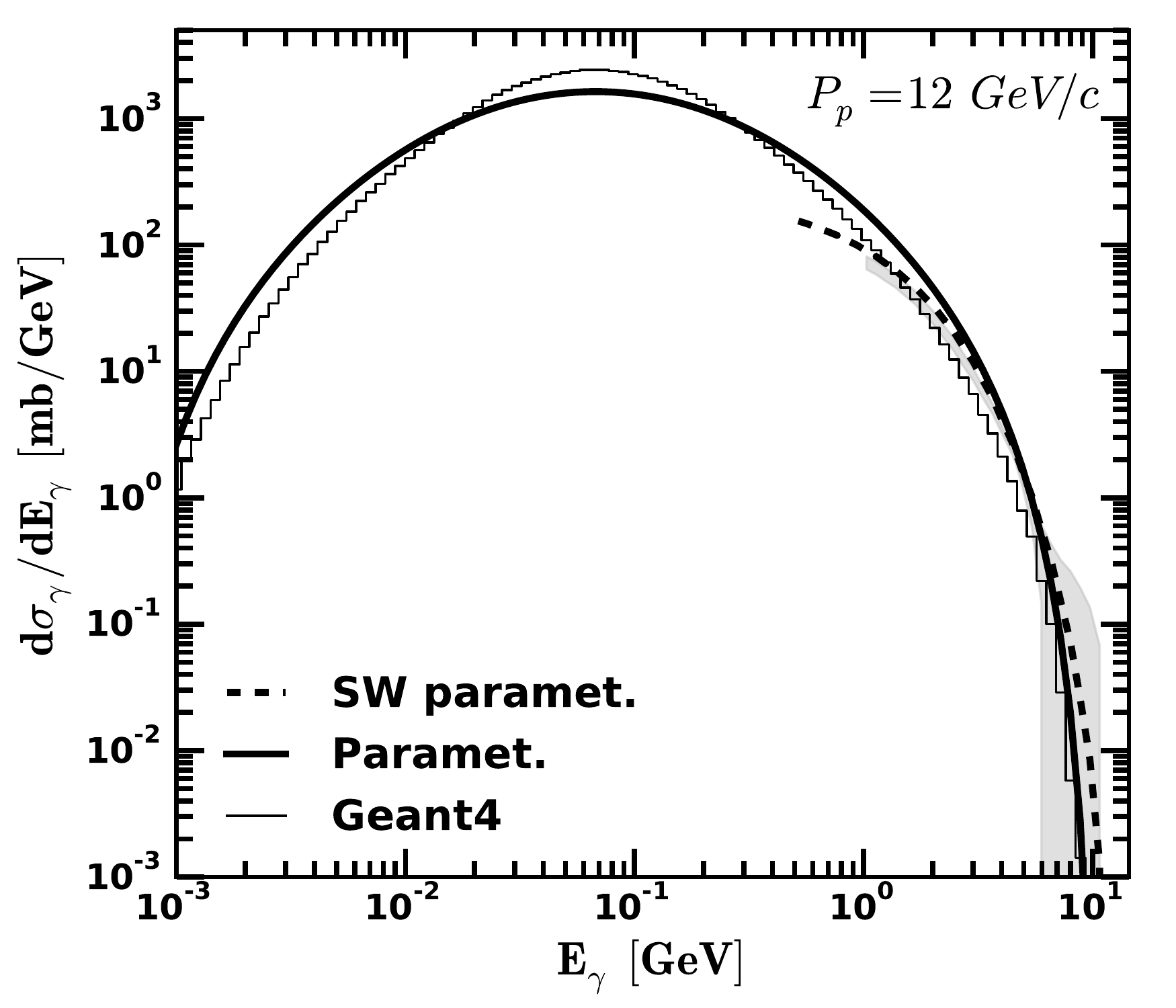}\\
\caption{The energy distribution of the $\pi^\pm$-mesons and their secondary products created from $p+C$ interactions at $P_p=3$, 5, 8, 12~GeV/c. Left column show the $\pi^\pm$ energy distribution and the data points are angle integrated HARP forward data \cite{Harp1,Harp2}. Middle column show the $e^\pm$ energy distribution from the $\pi^\pm$ decay and the right column show the $\gamma$-ray energy distribution from the $\pi^0$ decay. The $\pi^0$ differential cross section for the SW parametrization is obtained from the $\pi^\pm$ ones through the isospin relations betwee pion species. The histogram lines are the Geant4 predictions, the dash line are the predictions from the angle integrated Sanford-Wang (SW) parametrization \citep{Harp1} and the full line is the parametrization shown here. The black color in the left and middle columns show the $\pi^+$ and $e^+$ results, whereas, the gray represent $\pi^-$ and $e^-$ results. For visual effects, the $\pi^-$ results are divided by ten and the $e^-$ results are divided by one hundred. The gray areas represent the uncertainties of the $\gamma$-ray and $e^\pm$ production spectra when computed directly from the experimental pion data.\label{fig:HARP}}
\end{figure*}

\begin{figure*}
\includegraphics[scale=0.33]{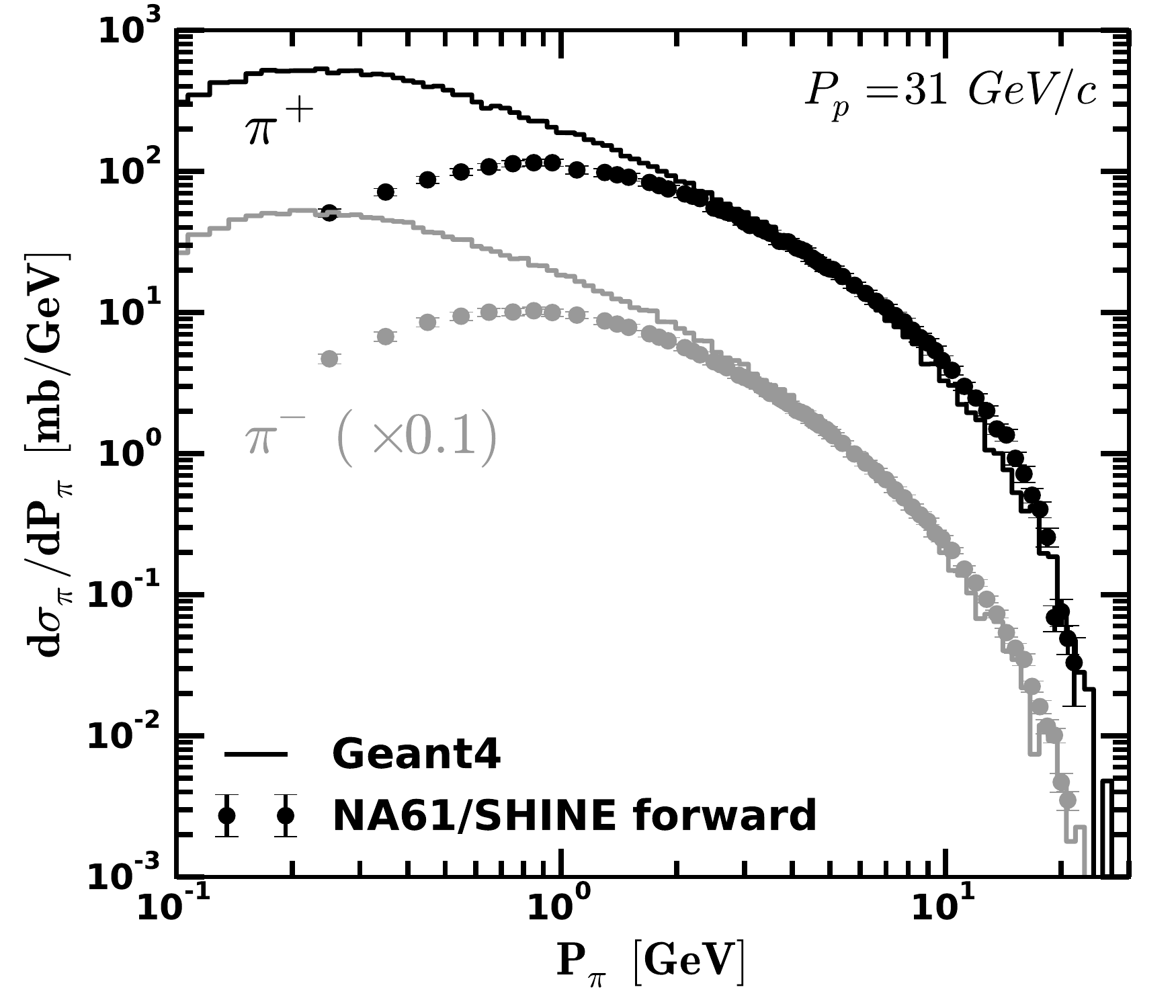}
\includegraphics[scale=0.33]{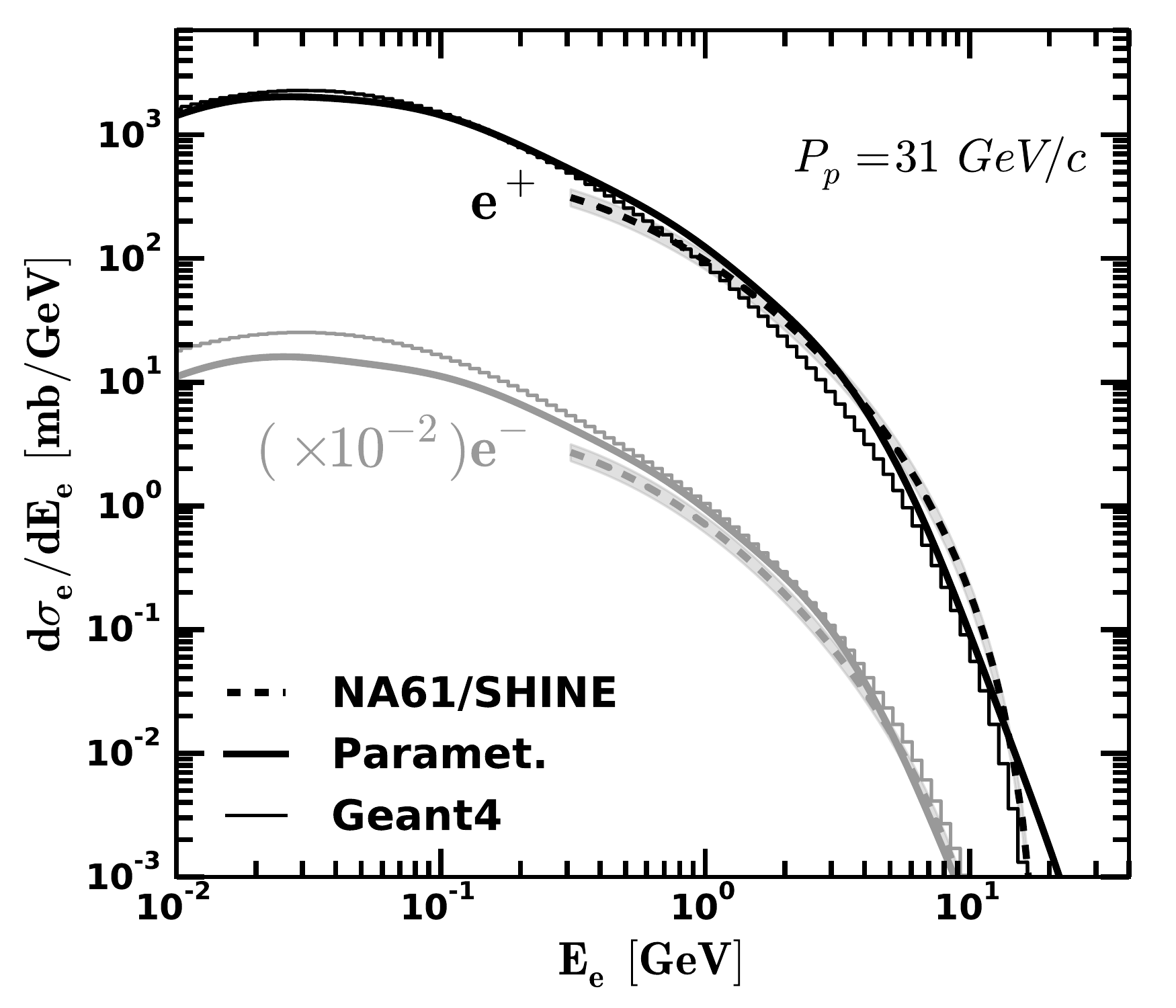}
\includegraphics[scale=0.33]{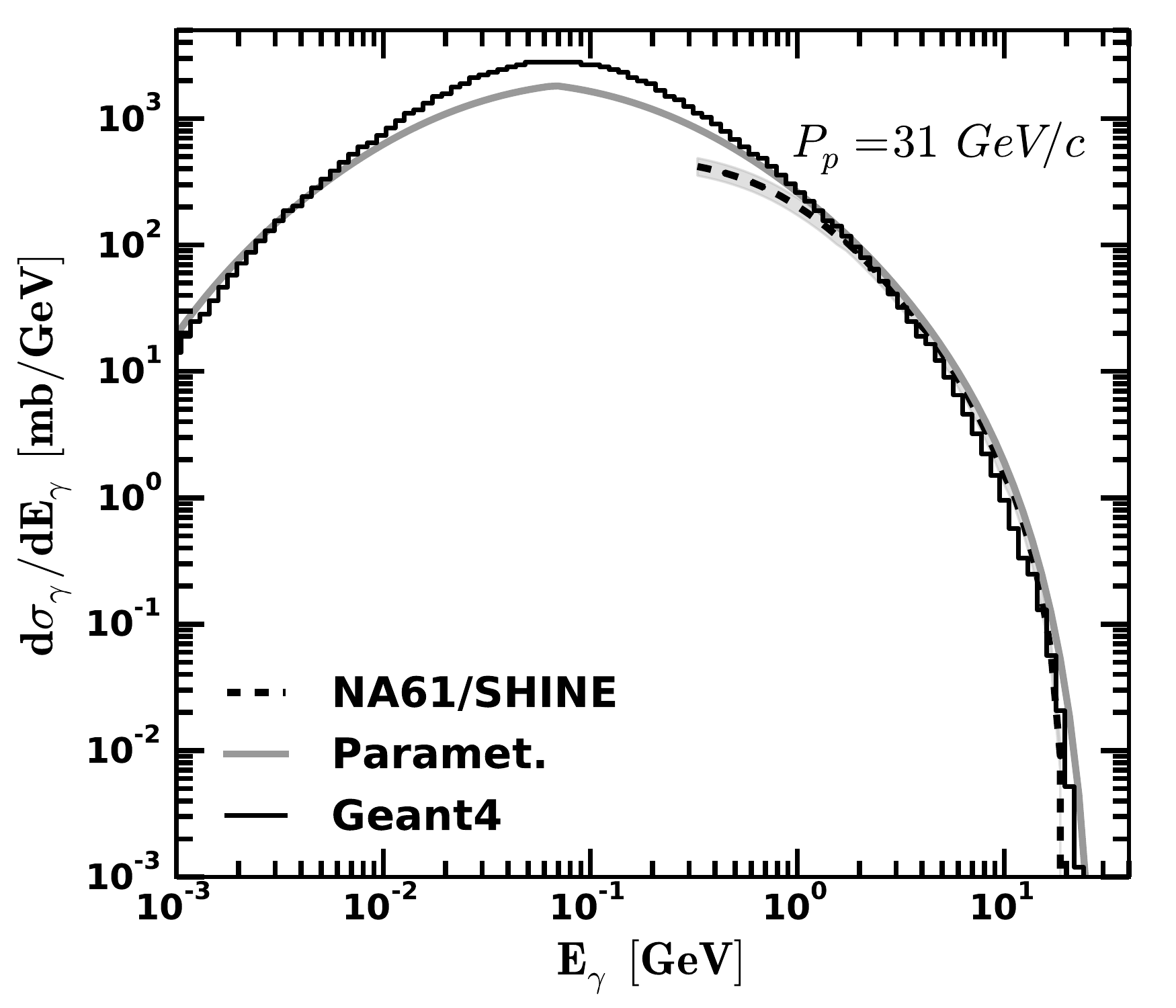}
\caption{The energy distribution of the $\pi^\pm$-mesons and their secondary products created from $p+C$ interactions at $P_p=31$~GeV/c. Figure on the left show the $\pi^\pm$ energy distribution and the data points are angle integrated NA61/SHINE forward data \cite{NA61a,NA61b}. Figure in the middle show the $e^\pm$ energy distribution from the $\pi^\pm$ decay and the figure on the right show the $\gamma$-ray energy distribution from the $\pi^0$ decay. The experimental $\pi^0$ differential cross section is obtained from the $\pi^\pm$ ones through the isospin relations. The histogram lines are the Geant4 predictions, the dash line are the predictions from fitting the $\pi^\pm$ forward data and the full line is the parametrization shown here. The black color on the left and middle figures show the $\pi^+$ and $e^+$ results, whereas, the gray represent the $\pi^-$ and $e^-$ results. For visual effects, the $\pi^-$ results are divided by ten and the $e^-$ results are divided by one hundred.\label{fig:NA61-SHINE}}
\end{figure*}

\begin{figure*}
\includegraphics[scale=0.33]{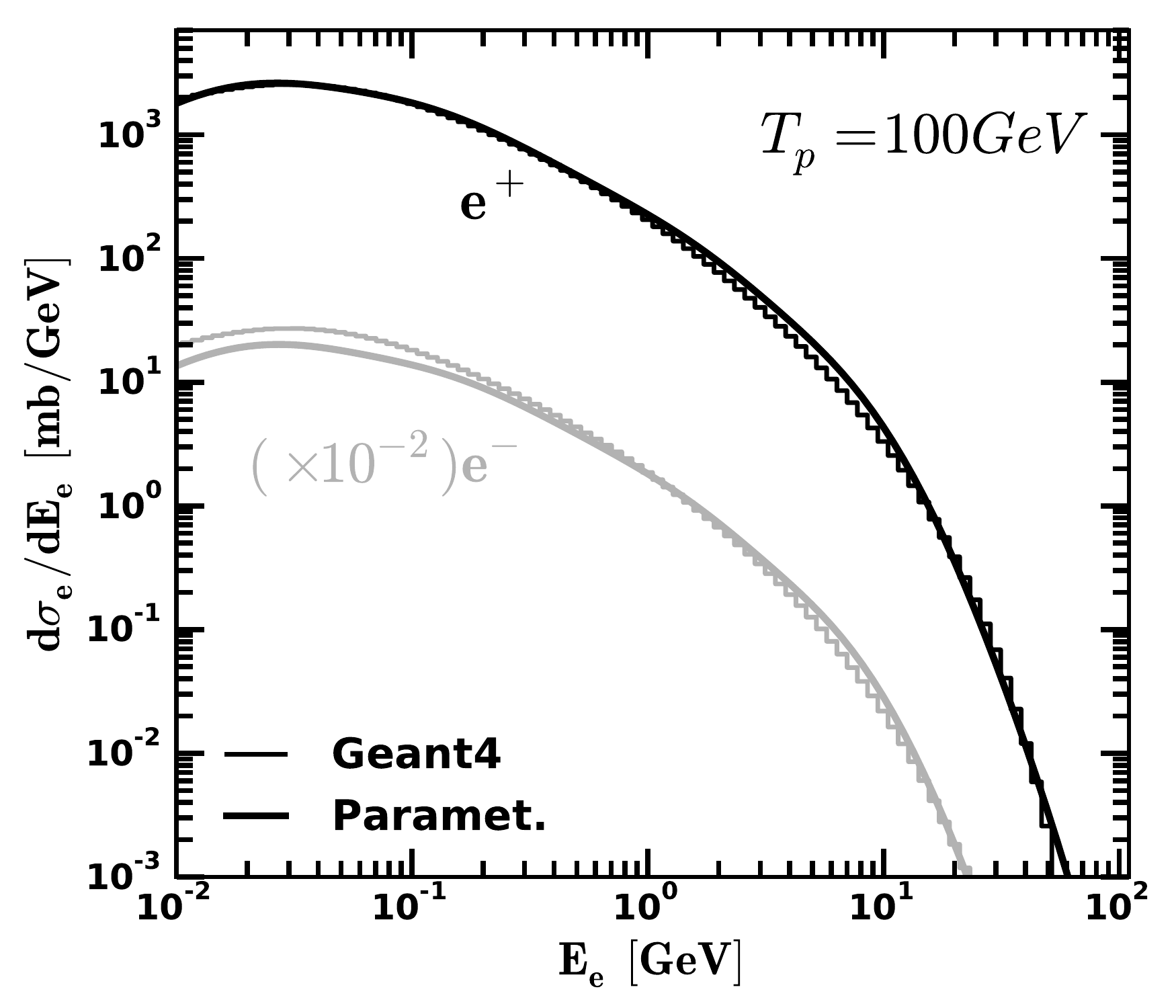}
\includegraphics[scale=0.33]{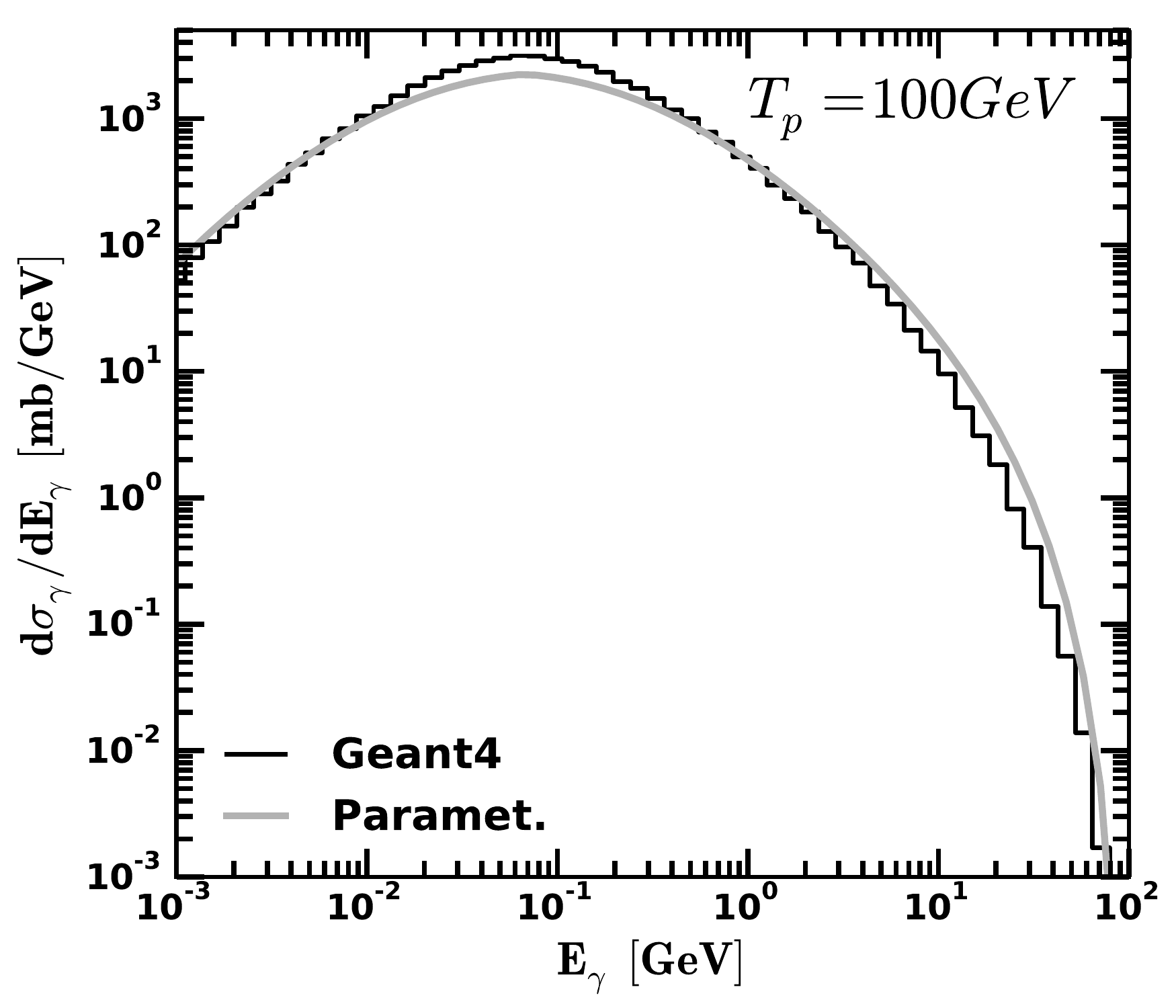}
\caption{Secondary particles energy distribution for $p+\rm{^{12}C}$ interaction at $T_p=100$~GeV. The histogram line represent the Geant4 results, whereas the full line is the parametrization presented here. The black lines in the left figure represent the $e^+$ results, whereas, the gray color represents the $e^-$ results that for visual effects are divided by one hundred. \label{fig:100GeV}}
\end{figure*}

For collision energies $1\leq T_p \leq 100$~A~GeV, the parametrization assumes similar secondary particle production spectra for $A+B$ and nucleon--nucleon interactions. To test this assumption Figs.~\ref{fig:HARP}--\ref{fig:100GeV} compare the results of the parametrization with the HARP and NA61/SHINE forward pion production experimental data \citep{Harp1, Harp2, NA61a, NA61b} and the results of the Geant4.10 Monte Carlo code \citep{Geant42003, Geant42006}. The hadronic model used in Geant4 simulations is the \textit{FTFP-BERT} which combines the Bertini intranuclear cascade model at low energies and the \textit{FRITIOF} string model at higher energies. These comparisons give a good estimate how well the parametrization describes the forward data and how well does it compare with more sophisticated hadronic monte carlo models.

Figures \ref{fig:HARP} and \ref{fig:NA61-SHINE} compare the energy distribution for: $p+C\to\pi^\pm$ (left column), $\pi^\pm\to e^\pm$ decay (middle column) and $\pi^0\to\gamma$-ray decay (right column) for collision momenta $P_p=3$, 5, 8, 12 and 31~GeV/c. The experimental data points for the $\pi^\pm$ energy distributions are obtained from the angular integration of the invariant cross section data \cite{Harp1, Harp2, NA61a, NA61b}. The histogram lines are the respective results from Geant4 simulations, whereas, the dash lines are the results of the angular integration of the Sanford--Wang (SW) parametrization of the forward pion production invariant cross section \citep{Harp1}. The dash line at $P_p=31$~GeV/c are the results of the direct fit of the experimental data. The gray areas represent the uncertainties of the $\gamma$-ray and $e^\pm$ production spectra when computed directly from the experimental pion data.

The left column of Figs~\ref{fig:HARP} and \ref{fig:NA61-SHINE} compare the forward $\pi^\pm$ spectral data with Geant4 predictions. It is clear from the figures that the Geant4 overestimates the forward pion production data for $P_p=3$ and 5~GeV/c and it agrees reasonably well for $P_p>5$~GeV/c. The SW parametrization also overestimates the $\pi^+$ production at $P_p=3$~GeV/c and it does not satisfy the kinematic limit, i.e. have non-zero cross sections for pion energy greater than the maximum allowed by the kinematics.

The full lines in the middle and right columns of Figs.~\ref{fig:HARP} and \ref{fig:NA61-SHINE} represent predictions from the parametrization developed here. The $\gamma$-ray energy distribution that is presented with dash lines, are computed from $\pi^\pm$ cross sections: 
\[\frac{d\sigma_{\pi^0}}{dE_{\pi^0}}=\frac{1}{2}\left(\frac{d\sigma_{\pi^+}}{dE_{\pi^+}} + \frac{d\sigma_{\pi^-}}{dE_{\pi^-}}\right),\]
where the $d\sigma_{\pi^\pm}/dE_{\pi^\pm}$ are the angle integrated forward data or the SW parametrization.

From the left column of Figs.~\ref{fig:HARP} and \ref{fig:NA61-SHINE}, we see that the forward experimental pion data agree with Geant4 predictions for $E_\pi>1$~GeV. This means that lower energy pion production is a result of large-angle emission. When comparing the $\gamma$-ray and $e^\pm$ energy distributions in the forward region above 1~GeV, good agreements are found between the experimental data plotted with the gray area and the SW parametrization, Geant4 and the parametrization developed here. The deviations are expected for energies below 1~GeV. We can see that the parametrization and the Geant4 predictions agree reasonably well for the $\pi^0\to2\gamma$ energy distribution. The maximum deviation is of the order 30--40~\%. For the $e^\pm$ energy distributions the deviations between these two models are larger. For $P_p\leq5$~GeV/c, the $e^-$ production spectra between these two models agree reasonably well, whereas, the $e^+$ spectra disagree by a factor as large as three. For $P_p>5$~GeV/c the $e^+$ spectra agrees reasonably good between these two models and disagree for the $e^-$ spectra by a factor as large as two. The root of these disagreements however, seem to be related with the total $e^\pm$ production yield. The parametrization developed here fits well the experimental pion yield data, thus, I believe that these large disagreements are not a result of the assumption of similar $e^\pm$ spectra between $p+$C and nucleon--nucleon.

As a last example, Fig.~\ref{fig:100GeV} compares the $p+C$ secondary particle production at $T_p=100$~GeV between Geant4 and the parametrization introduced here. These two models agree reasonably well at this collision energy with differences less than 35~\%.

\section{Results and Discussion}
To illustrate the results of the parametrizations developed here, let us consider two examples. The first one considers $p$ and $^{12}$C projectiles interacting with the $^{12}$C target material. The projectile fluxes are described in one case by a power-law function with index $\alpha=3$ and, in the second case by the same power-law but with an exponential cut-off at $T_p^{\rm cut}=1$~A~GeV. In the second example, the chemical composition of both projectiles and target material are considered to be similar to the solar composition of elements. The projectile fluxes for this example are assumed to be in one case a pure power-law function with index varying between $2\leq \alpha \leq 5$, whereas, in the second case the same power-law function but with an exponential cut-off at $T_p^{\rm cut}=1$~A~GeV.

Consider the following functional form for the projectile flux:
\begin{equation}\label{eq:JA}
J_A(T_p)=\mathcal{N}\times T_p^{-\alpha}\times\exp\left(-\frac{T_p}{T_p^{\rm cut}} \right).
\end{equation}
\noindent Assuming that the target number density is $n_B$, the $\gamma$-ray production spectrum is given by:
\begin{equation}\label{eq:prodGam}
\frac{dN}{dE_\gamma}=4\pi\,n_B \int\limits_{T_p^{\rm th}}^\infty dT_p\;J_A(T_p)\; \frac{d\sigma_\gamma^{AB}}{dE_\gamma}(T_p,E_\gamma).
\end{equation}
\noindent Here, $d\sigma_\gamma^{AB}/dE_\gamma$ is the sum of the $\gamma$-ray production cross sections via formation of hard photons ($A+B\to\gamma$) and decay of secondary $\pi^0$-mesons ($A+B\to\pi^0$). $T_p^{\rm th}$ is the kinematic threshold kinetic energy per nucleon. For $\pi^0$ production $T_p^{\rm th}$ is given by Eq.~(\ref{eq:Tpth}). For producing hard photons with $E_\gamma\geq E_\gamma^{\rm min}=30$~MeV it is given by:
\[T_p^{\rm th}=\left(\frac{1}{A_p}+\frac{1}{A_t}\right)E_\gamma^{\rm min},\]

\noindent where, $A_p$ and $A_t$ are the mass numbers for the projectile $A$ and the target $B$, respectively. The $e^\pm$ pair production spectra is calculated using Eqs.~(\ref{eq:dNdEpiepm}, \ref{eq:dNdEe} and \ref{eq:JA}).

Calculation of the $\gamma$-ray spectrum for a solar-like composition of elements, includes a sum over all possible nuclear collisions. Let $n_B$ be the element $B$ number density in the target material and $J_A$ be the projectile $A$ flux defined in Eq.~(\ref{eq:JA}). In addition, let us assume that all projectiles fluxes have the same $\alpha$ and $T_p^{\rm cut}$ parameters. If we note with $X_p^A=J_A/J_p$ the flux ratio between the element $A$ and the proton, and $X_t^B=n_B/n_H$ the target number density ratio between the element $B$ and hydrogen, then the Eq.~(\ref{eq:prodGam}) transforms to: 
\begin{equation}\label{eq:SolarprodGam}
\frac{dN}{dE_\gamma}=\mathcal{C}\sum\limits_{A,B} X_p^A\,X_t^B \!\!\!\!\!\!\int\limits_{T_p^{\rm th}(AB)}^\infty \!\!\!\!\!\!\!\! dT_p\;J(T_p) \frac{d\sigma_\gamma^{AB}}{dE_\gamma}(T_p,E_\gamma).
\end{equation}
\noindent Here, $T_p^{\rm th}(AB)$ is the threshold energy for the specific $A+B$ interaction, $J=J_A/\mathcal{N}_A$ is the same for all projectiles, and $\mathcal{C}=4\pi \,n_p\,\mathcal{N}_p$ is a normalization constant. When using this formula it is important to recall that the hard photon cross section is calculated for projectiles lighter or equal the mass of the target nucleus. Therefore, the contribution of $A+B$, when $A$ is heavier than $B$, is calculated from $B+A$. 

For a solar-like composition of elements: H, $^4$He, $^{12}$C, $^{14}$N, $^{16}$O, $^{20}$Ne, $^{24}$Mg, $^{28}$Si, $^{32}$S and $^{56}$Fe, we have $X_p=X_t=1:9.59\times10^{-2}: 4.65\times10^{-4} :8.3\times 10^{-5}: 8.3 \times 10^{-4} :1.2 \times 10^{-4} :3.87 \times 10^{-5} :3.69 \times 10^{-5}: 1.59 \times 10^{-5} :3.25 \times 10^{-5}$ \citep[see e.g.][]{Meyer1985}. 

At higher energies nucleus--nucleus and $pp$ $\gamma$-ray spectral shape are similar, therefore, by using the wounded nucleon model \citep{Bialas1976} we can scale the $\gamma$-ray spectra from $pp$ interactions with the following factor:
\begin{equation}
\epsilon=\sum\limits_{A,B} X_p^A\,X_t^B\,\frac{\sigma_R(AB)\, W_{AB}}{2\,\sigma_{pp}}.
\end{equation}
\noindent Where, $\sigma_R(AB)$ is the reaction cross section for $A+B$ interactions, see Eq.~(\ref{eq:XSr}). $W_{AB}=(A_p\,\sigma_{pB}+A_t\,\sigma_{pA})/\sigma_{AB}$ is the number of wounded nucleons and $\sigma_{pA}$, $\sigma_{pB}$ and $\sigma_{AB}$ are the inelastic $p+A$, $p+B$ and $A+B$ cross sections calculated using parametrization \citep{Shen1989}. The $\sigma_{pp}$ is the $pp$ inelastic cross section and is taken from \citep{Kafexhiu2014}.

\begin{figure*}
\includegraphics[scale=0.43]{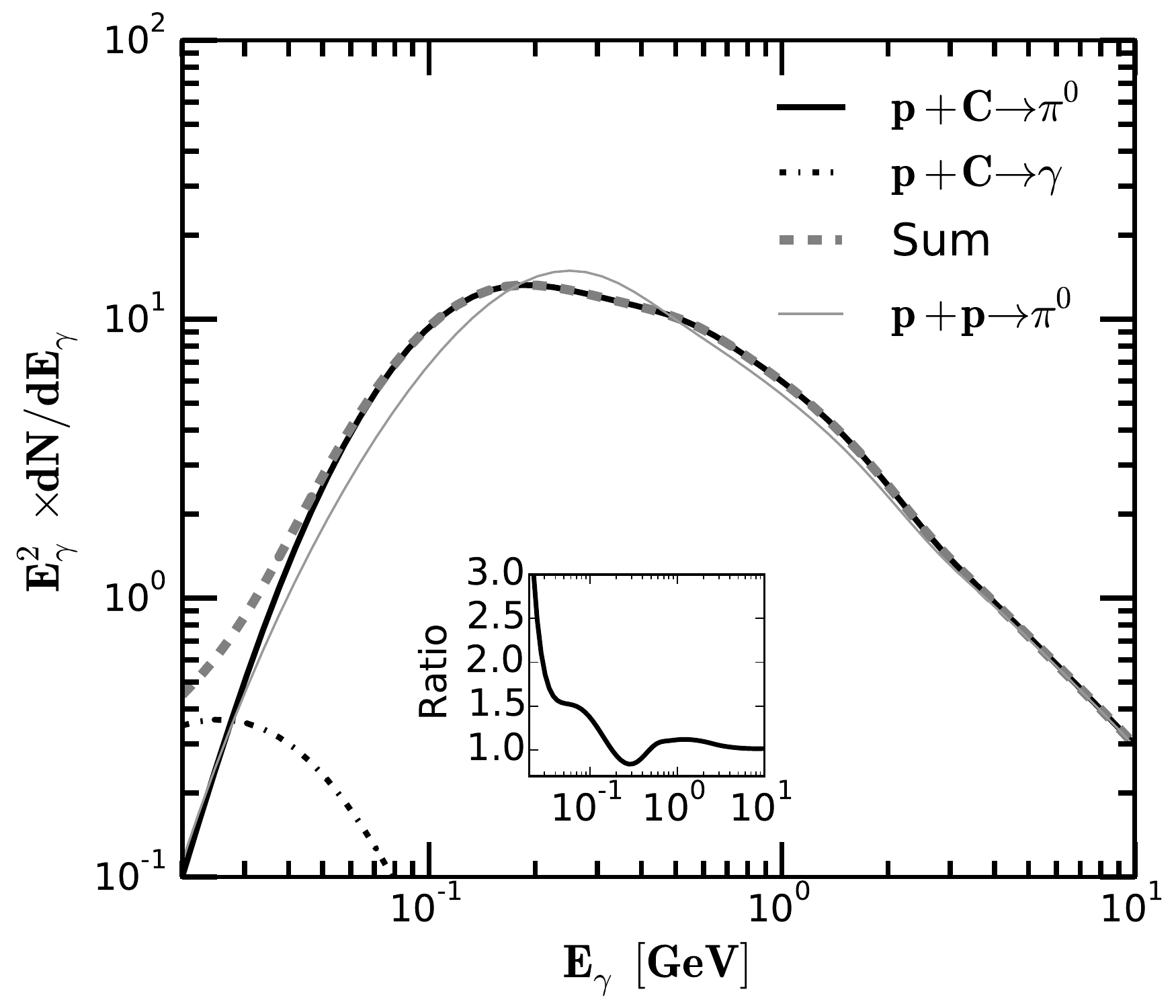}
\includegraphics[scale=0.43]{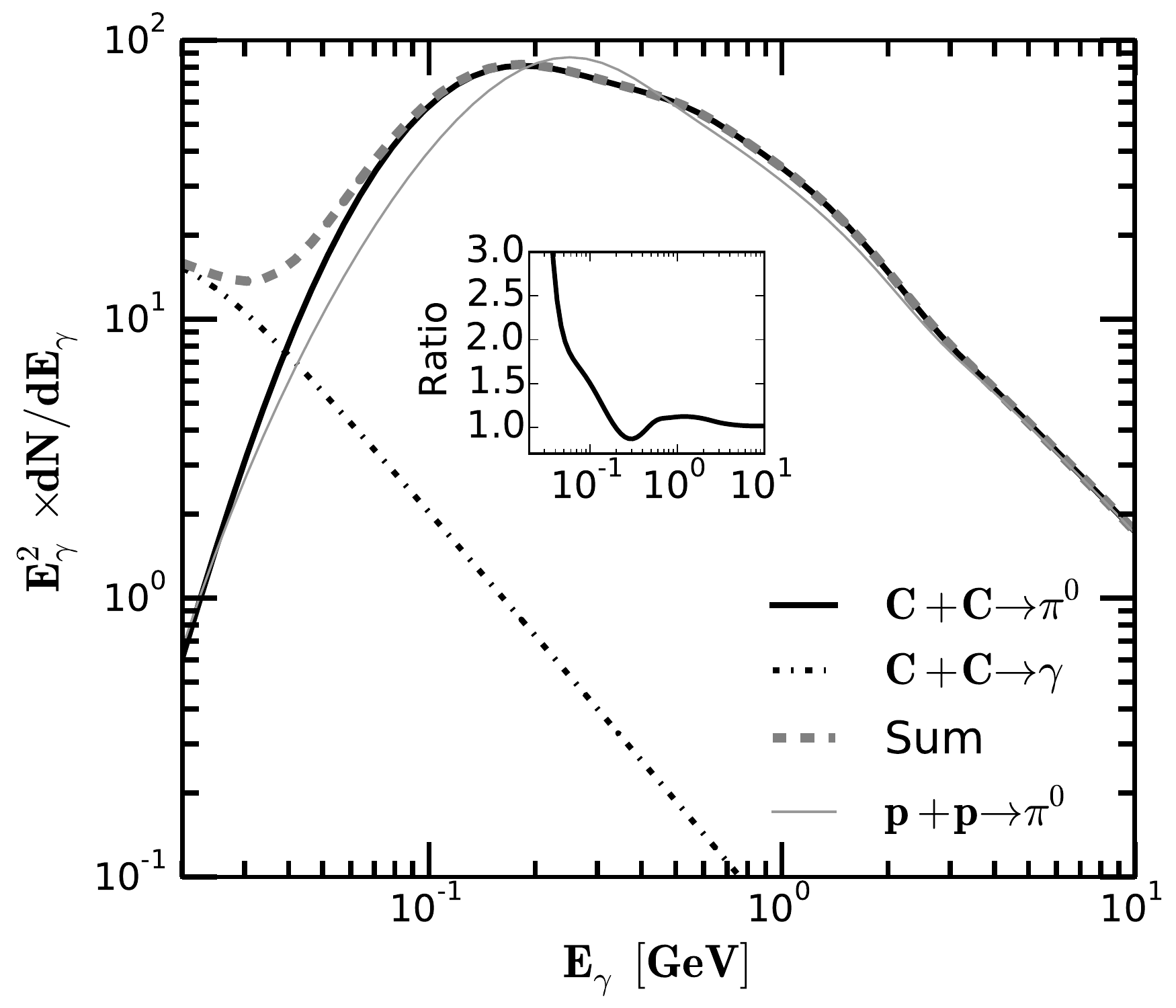}\\
\includegraphics[scale=0.43]{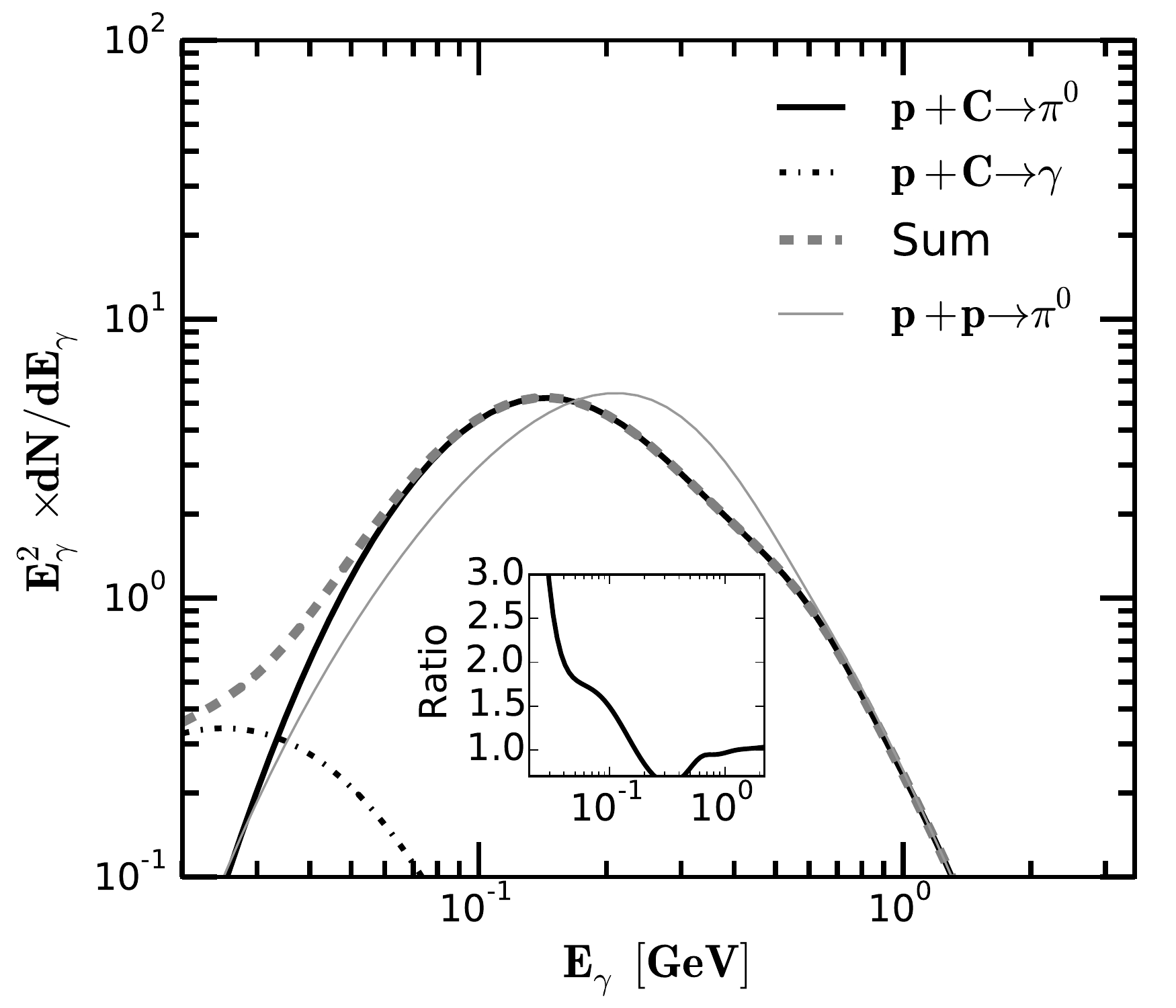}
\includegraphics[scale=0.43]{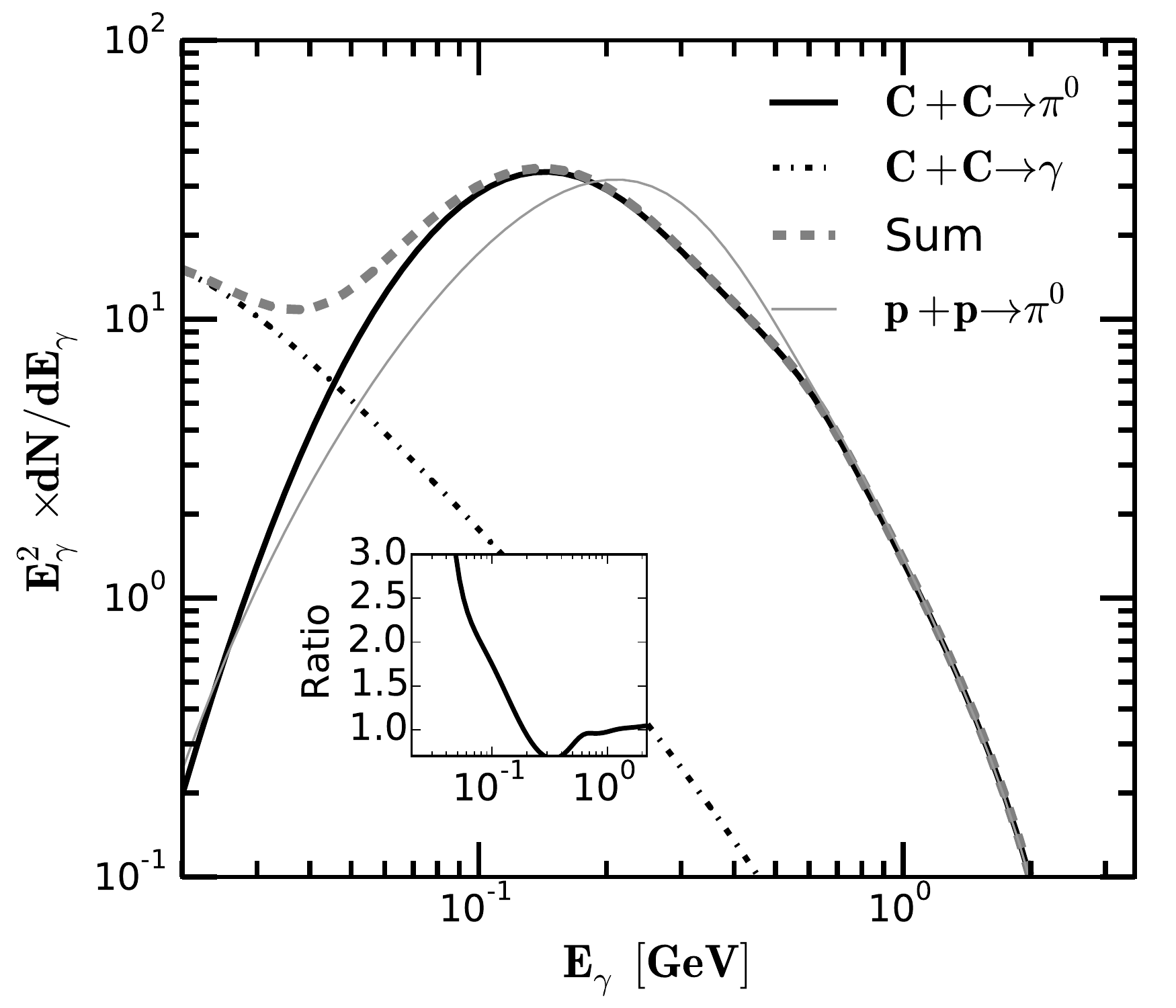}
\caption{Gamma-ray production spectra for $p+^{12}$C (left) and $^{12}$C+$^{12}$C (right) interactions. The projectile fluxes have a power-law index $\alpha=3$ and in one case $T_p^{\rm cut}\to \infty$ (top) and in the other case $T_p^{\rm cut}=1$~A~GeV (bottom), see Eq.~(\ref{eq:JA}). The thick black line is the $\pi^0\to2\gamma$ contribution, doted line is the hard photon contribution and the dash line is the sum. The thin gray line correspond to $pp\to\pi^0$ scaled according to the wounded nucleon model. The small panel in each plot show the spectral ratio between the nucleus--nucleus and $pp$. \label{fig:GamSpecPow}}
\end{figure*}

\begin{figure*}
\includegraphics[scale=0.43]{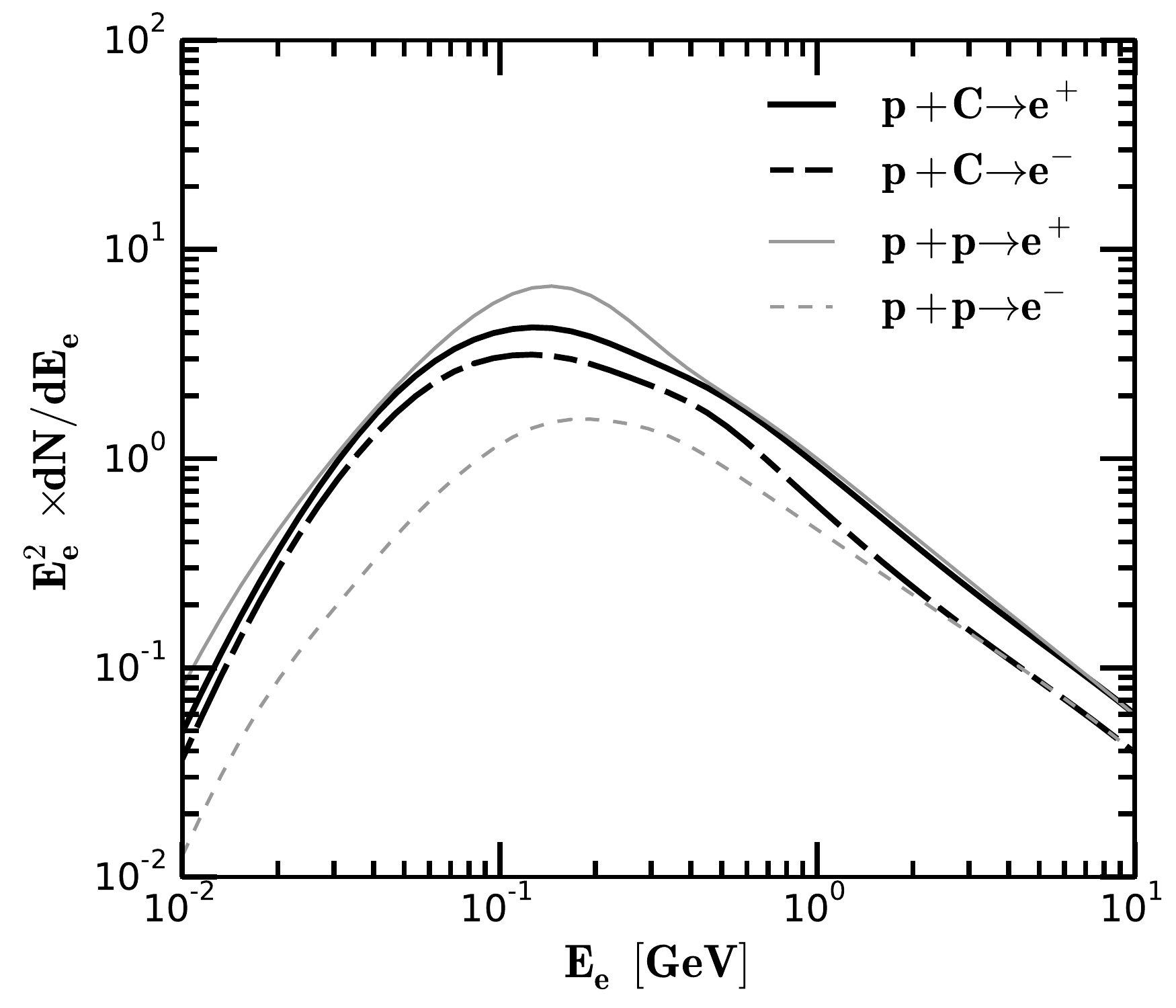}
\includegraphics[scale=0.43]{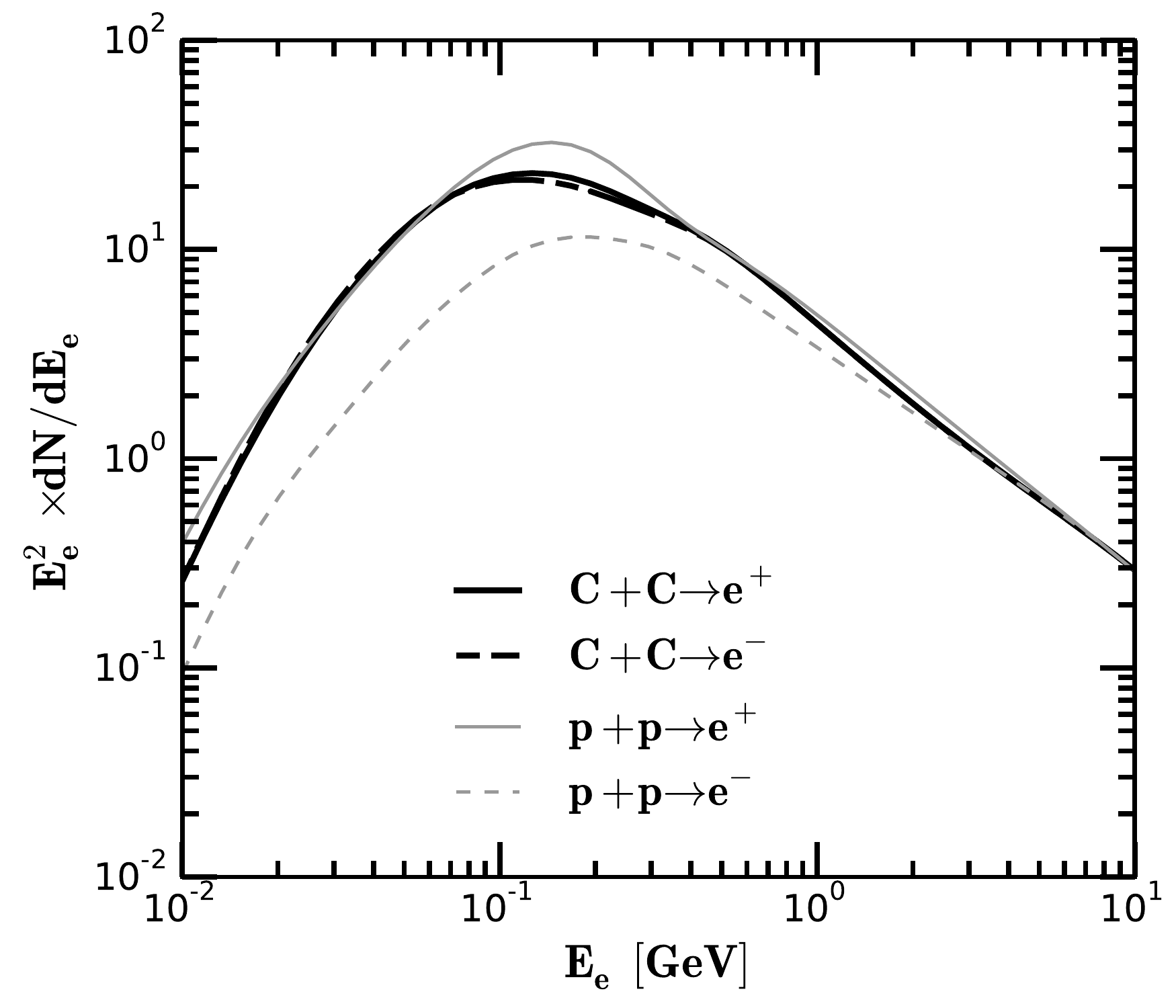}\\
\includegraphics[scale=0.43]{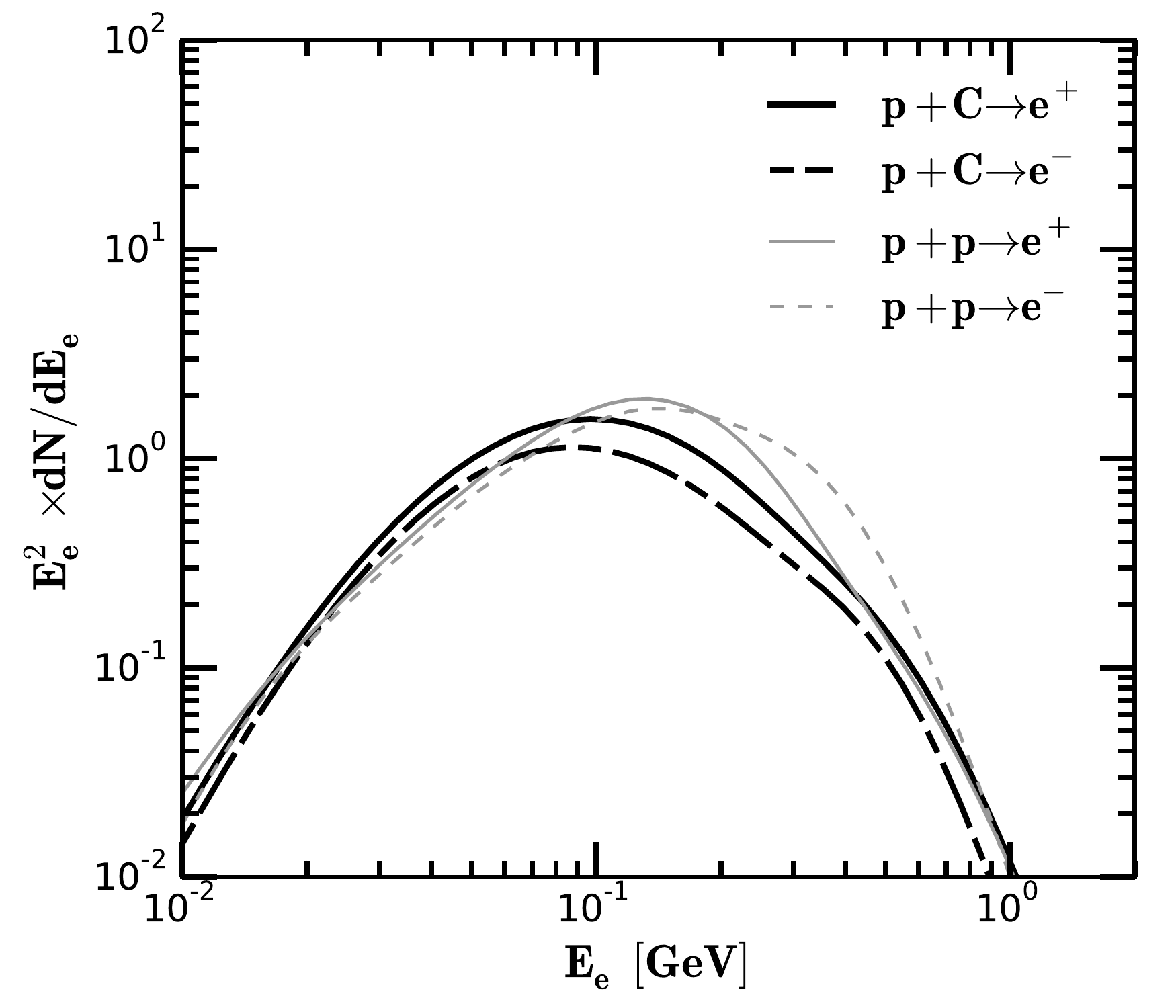}
\includegraphics[scale=0.43]{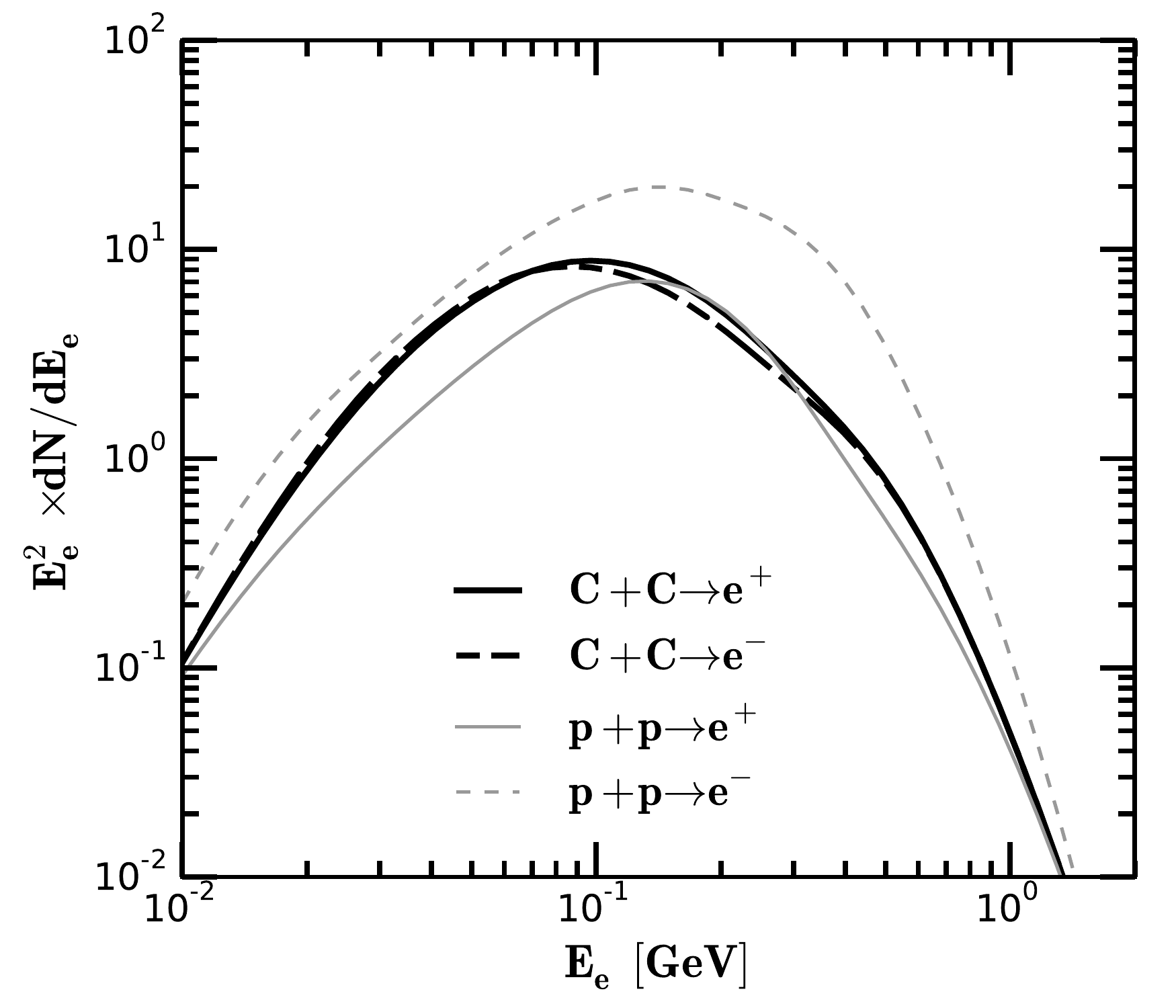}
\caption{Electron and positron production spectra for $p+^{12}$C (left) and $^{12}$C+$^{12}$C (right) interactions. The projectile fluxes have a power-law index $\alpha=3$ and in one case $T_p^{\rm cut}\to \infty$ (top) and in the other case $T_p^{\rm cut}=1$~A~GeV (bottom), see Eq.~(\ref{eq:JA}). The thick black line corresponds to $e^+$ production and the thick dash-line correspond to $e^-$ production. The thin gray line shows the contribution from $pp\to e^+$ and the thin dash-doted line shows the contribution from $pp\to e^-$ which are scaled for each case according to the wounded nucleon model. \label{fig:epSpecPow}}
\end{figure*}

\begin{figure*}
\includegraphics[scale=0.46]{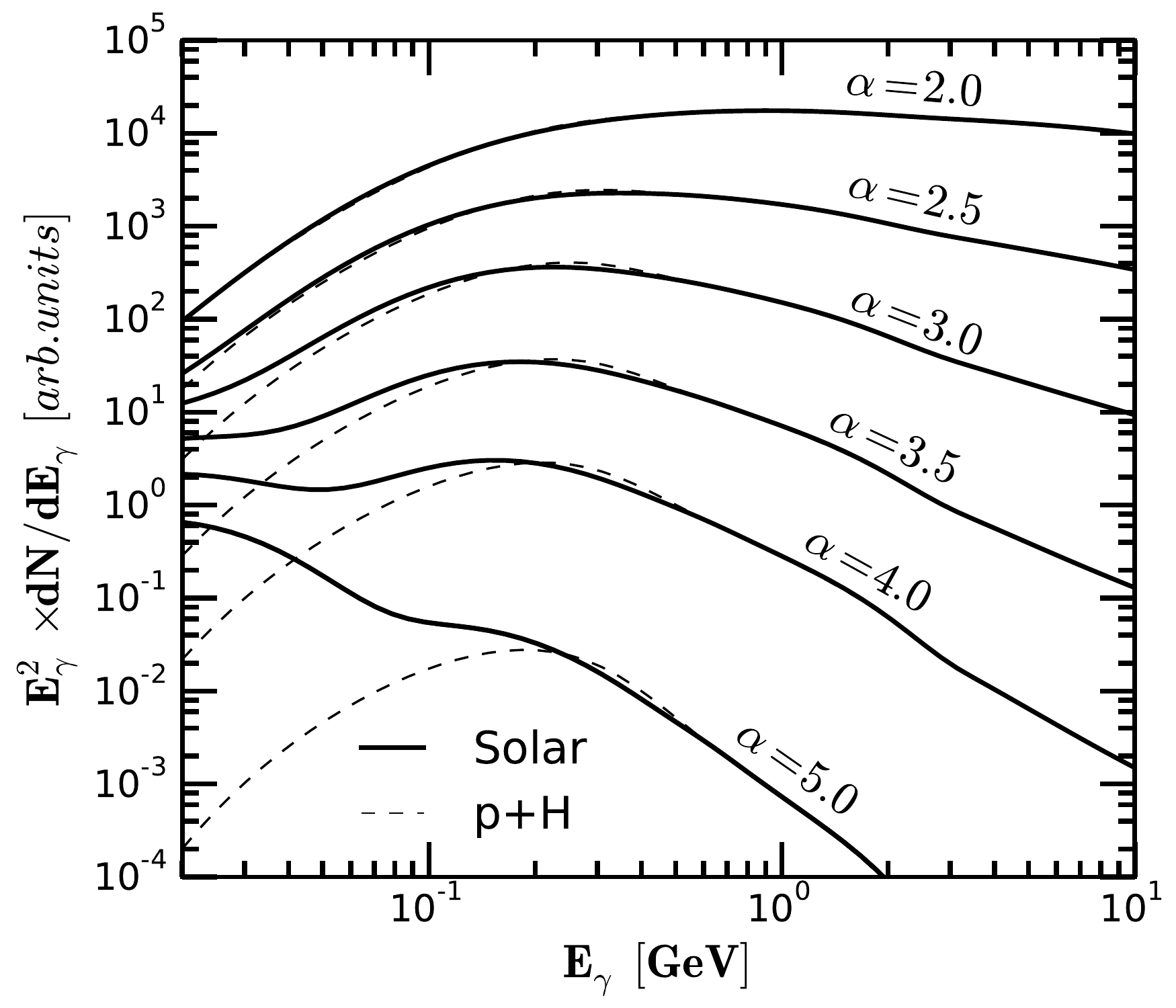}
\includegraphics[scale=0.46]{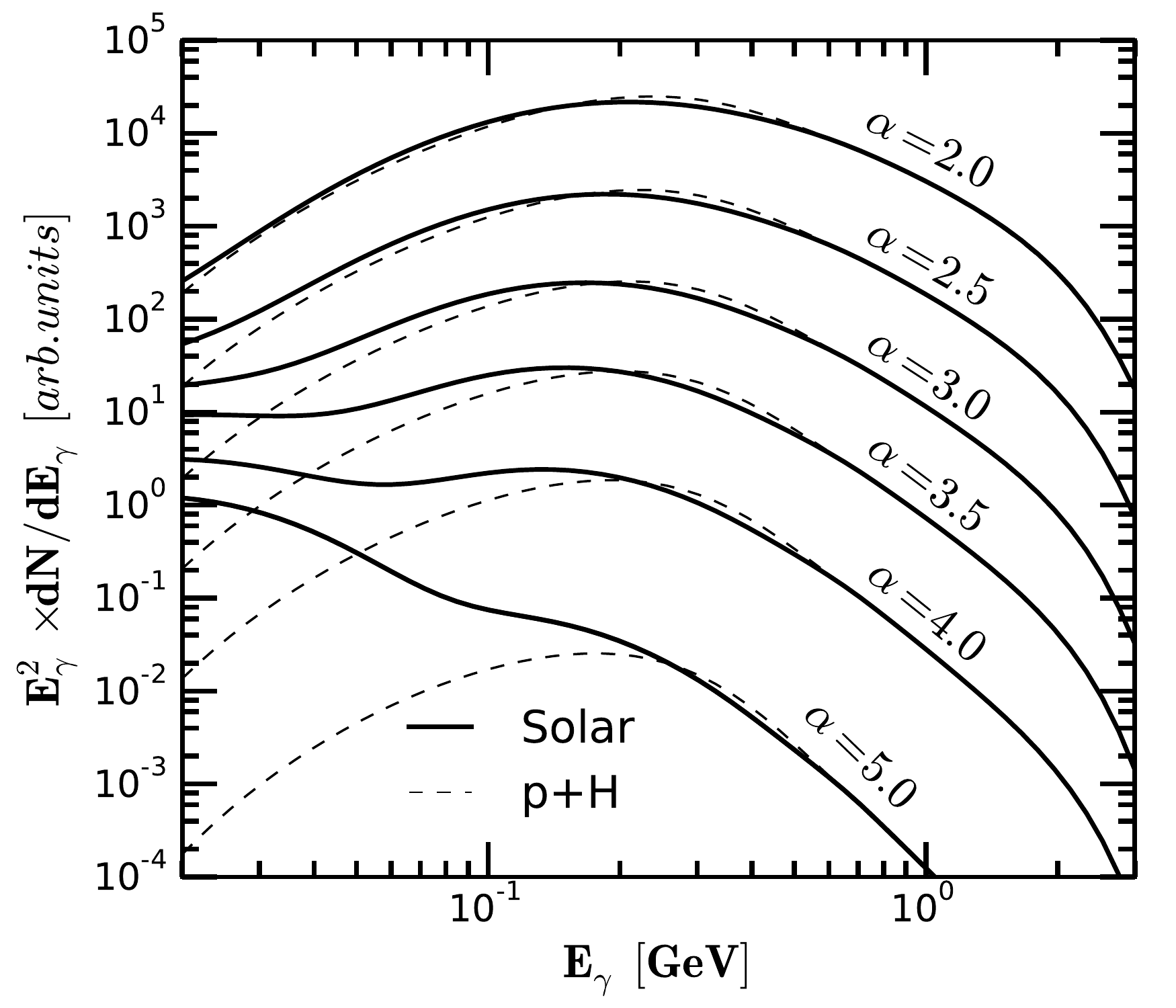}
\caption{Gamma-ray production spectrum from the hard photons and $\pi^0$-meson decay for a solar-like composition of elements. The exponential cut-off energies are $T_p^{\rm cut}\to \infty$ (left) and $T_p^{\rm cut}=1$~A~GeV (right), whereas, the power-law index varies between $2\leq\alpha\leq 5$. The solid lines represent the contribution from all nuclei, whereas, the thin dash lines represent the $\gamma$-ray spectra from $pp$ collisions which are scaled according to the wounded nucleon model. \label{fig:Solar}}
\end{figure*}

Figures~\ref{fig:GamSpecPow} and \ref{fig:epSpecPow} show the $\gamma$-ray and $e^\pm$ spectra resulting from $p+^{12}$C and $^{12}$C+$^{12}$C interactions. Looking at these figures, one can draw some general conclusions that are related to subthreshold pions and hard photon production: The $pp$ interactions fail to reproduce the $A+B$ secondary particle shape below 1~GeV. The $A+B$ interactions produce a secondary particle spectra that is generally broader than the respective $pp$ one. Direct hard photons manifest themselves for $E_\gamma<100$~MeV and this component produce a unique feature in the final $\gamma$-ray spectrum that has no analogue in $pp$ interactions. Finally, the differences between $A+B$ and $pp$ secondary particle production spectra depends on the nuclear mass and it increases when the masses increase.

The presence of a low energy exponential cut-off suppresses the high energy collisions. As a result, the relative contribution of hard photons and subthreshold pions increases. The low energy cut-off creates two effects: Firstly, it increases the relative $\gamma$-ray production rate for $E_\gamma<200$~MeV, which is unique for nuclear interactions, see Fig~\ref{fig:GamSpecPow}. Secondly, the peak of the $\gamma$-ray and $e^\pm$ spectra produced from nuclear interactions shifts toward lower energies at a higher rate compared to $pp$ interactions. For the $\alpha=3$ and $T_p^{\rm cut}=1$~A~GeV example that was considered here, the peak energy of the secondary spectra computed from nuclear collisions is about 40~\% smaller than that of the $pp$, see Fig.~\ref{fig:GamSpecPow} and \ref{fig:epSpecPow}. 

Independent of the projectile flux, the $\gamma$-ray spectrum for $E_\gamma<100$~MeV between $p+^{12}{\rm C}$ and $^{12}$C+$^{12}{\rm C}$ is different. This is due to different hard photon production cross sections between these two nuclear interactions. The $^{12}$C+$^{12}{\rm C}$ hard photon production cross section is larger and its spectrum is harder compared to $p+^{12}{\rm C}$. Therefore, the shape of the spectrum for $E_\gamma<100$~MeV can be used to discriminate not only $pp$ from nuclear collisions but also between different nucleus--nucleus interactions and it can be used in this way to estimate the mass of the colliding nuclei.

Proton--proton interactions produce $\pi^+$ for collision energies $T_p>0.28$~GeV and $\pi^-$ for $T_p>1.2$~GeV. Nucleus--nucleus interactions on the other hand, produce $\pi^\pm$ at much lower energies and with equal amount. This is a consequence of the isospin symmetry and having equal number of protons and neutrons in both projectile and target nuclei. Therefore, nuclear interactions produce similar amounts of $e^+$ and $e^-$ with similar spectral shape, in contrast to $pp$ interactions, which produce an excess of $e^+$ compared to $e^-$ at low energies. The exponential cut-off at $T_p^{\rm cut}=1$~A~GeV suppresses the $p+p\to e^-$ production and because of this, its shape is very different when compared with $A+B\to e^-$ spectrum, see Figure~\ref{fig:epSpecPow}. For $p+^{12}$C interaction, the projectile has no neutrons and as a result, the amount of $\pi^+$ ($e^+$) would be slightly larger than $\pi^-$ ($e^-$).

Figure~\ref{fig:Solar} shows the results for a solar composition of elements. The discussion about $p+^{12}{\rm C}$ and $^{12}$C+$^{12}{\rm C}$ interactions is also valid here. Furthermore, these calculations show that high energy nuclear interactions that result from hard spectrum of projectiles, produce abundant low energy $\gamma$-rays that screen the contribution of subthreshold pions and hard photons. Thus, the $\gamma$-ray spectrum from $A+B$ collisions with a primary spectrum harder than $\alpha\lesssim 2.5$, show no difference from $pp$. On the other hand, softer primary spectra and low energy cut-off produce significant differences between $A+B$ and $pp$ final $\gamma$-ray spectra below 200~MeV. This can be important in many astrophysical situations such as solar flares and oxygen-rich supernova remnants where soft ion spectra are produced and one cannot extrapolate $pp$ calculations for $E_\gamma\leq 200$~MeV.

\section{Summary and Conclusions \label{sec:conclusion}}

In this work, using numerous publicly available experimental data sets, simple and reasonably accurate parametrization formulas are provided to calculate the $\gamma$-ray and $e^\pm$ pair production spectra for nucleus--nucleus interactions with collision energy $T_p\leq 100$~A~GeV including the low energy subthreshold pion and hard photon channels. The parametrizations are in reasonably good agreement with the available experimental data and the expected deviations are less than 30--40~\%. These parametrizations are provided in the form of a computer library in \citep{LibNucNucGam}.

I conclude by stressing the importance of the subthreshold pion and hard photon channels for computing the secondary particle production spectra from low energy nuclear collisions. The $\gamma$-ray spectra produced by these two channels below 200~MeV cannot be reproduced by $pp$ interactions. This characteristic feature can be used to experimentally distinguish the contribution of nuclei from protons. While the detection of $\gamma$-rays for $E_\gamma \gtrsim 1$~GeV can be used to fix the primary spectra parameters, the detection of $E_\gamma \lesssim 200$~MeV can be used to estimate the masses of the colliding nuclei. Gamma-ray instruments, such as Fermi-LAT, that are sensitive in this energy region should include the contribution of these two channels in their $\gamma$-ray analysis especially for astrophysical sources that have soft primary nuclear spectra. 

Future work will improve the accuracy of the pion energy distribution. This is possible if at low energies a two temperature pion spectra is considered instead of one. At high energies one could parametrize the pion spectrum directly from the proton--nucleus and nucleus--nucleus experimental data.

\begin{acknowledgments}
The author thanks the MPIK and its High Energy Astrophysics Theory Group
for their support. He would like to thank Felix Aharonian for fruitful and motivating discussions as well as Andrew M. Taylor and Roland Crocker for their helpful comments to improve the text.
\end{acknowledgments}

\appendix 
\section{Calculation of the pion energy distribution using $pp$ cross sections \label{append:A}}
High energy $A+B$ inelastic collisions produce pions through $pp$, $np$ and $nn$ interactions. The average pion yield from these individual interactions are different, therefore, a correct estimation of the pion production cross section from nucleus--nucleus collisions should average over different nucleon--nucleon contributions. Let the $Z_p$, $A_p$ and $Z_t$ and $A_t$ be the number of protons and the total number of nucleons for the projectile $A$ and the target $B$. Let us define $\xi_p=Z_p/A_p$ and $\xi_t=Z_t/A_t$ the probabilities of randomly colliding with a proton inside the projectile and the target nuclei, respectively. The quantities $1-\xi_p$ and $1-\xi_t$ express the probabilities for colliding with a neutron. Using these probabilities, the average  pion multiplicity per nucleon--nucleon collision is:
\begin{equation}\label{eq:NpiAB}
\begin{split}
\left\langle \pi_{NN}\right\rangle=\xi_p\xi_t\left\langle \pi_{pp}\right\rangle \! &+\! [\xi_p(1-\xi_t) + \xi_t(1-\xi_p)]\left\langle \pi_{np}\right\rangle\\ &+ (1-\xi_p)(1-\xi_t)\left\langle \pi_{nn}\right\rangle,
\end{split}
\end{equation}

\noindent where, $\left\langle \pi_{pp}\right\rangle$, $\left\langle \pi_{np}\right\rangle$ and $\left\langle \pi_{nn}\right\rangle$ are the pion average production multiplicities for $pp$, $np$ and $nn$ collisions, respectively. The multiplicity for $A+B$ collisions is $\left\langle \pi_{AB}\right\rangle \sim \left\langle \pi_{NN}\right\rangle$ and the proportionality factor is the number of participating nucleons. 

Assuming that charge and isospin symmetries hold at $T_p>1$~A~GeV, then the average pion production multiplicities satisfy the following relations \citep[see e.g.][]{Golokhvastov2001a}:
\begin{equation}\label{eq:Npi_relation}
\begin{array}{l}
\left\langle \pi_{pp}^-\right\rangle = \left\langle  \pi_{nn}^+\right\rangle \\
\left\langle\pi_{pp}^0\right\rangle = \left\langle\pi_{nn}^0\right\rangle = \left\langle \pi_{np}^{\pm,0}\right\rangle \\
\left\langle \pi_{pp}^+\right\rangle=\left\langle\pi_{nn}^-\right\rangle \\
\end{array}
\end{equation}

From these relations follows that $\left\langle\pi_{pp}^0\right\rangle=\left(\left\langle \pi_{pp}^+\right\rangle + \left\langle \pi_{pp}^-\right\rangle\right)/2$ which is used in the text. Moreover, assuming that these relations extent to differential cross sections, then, by manipulating Eqs.~(\ref{eq:NpiAB} and \ref{eq:Npi_relation}) follows that:

\begin{equation}\label{eq:dXSdEpiMultip}
\begin{array}{l}
f_{AB}^{\pi^+} = \left(\frac{\xi_p+\xi_t}{2}\right)\times f_{pp}^{\pi^+} + \left(1-\frac{\xi_p+\xi_t}{2}\right)\times f_{pp}^{\pi^-},\\ 
\\
f_{AB}^{\pi^0} = f_{pp}^{\pi^0},\\ 
\\
f_{AB}^{\pi^-} = \left(1-\frac{\xi_p+\xi_t}{2}\right)\times f_{pp}^{\pi^+} + \left(\frac{\xi_p+\xi_t}{2}\right)\times f_{pp}^{\pi^-},\\ 
\end{array}
\end{equation}

\noindent where, $f=\sigma_{\rm inel}^{-1}\times d\sigma/dE_\pi$ is the pion energy distribution for $A+B$ and $pp$ interactions. These relations are used to calculate the $A+B\to (e^\pm,~\gamma)$ energy distribution using $p+p\to (e^\pm,~\gamma)$ parametrizations.

\bibliographystyle{apsrev}
\bibliography{SubthreshPi0,Brems,auxillary}
\end{document}